\documentstyle[amssymb,preprint,aps,rmp]{revtex}

\begin{document}
\title{\smallskip A genuine reinterpretation of the Heisenberg's (``uncertainty'')
relations }
\author{Spiridon Dumitru}
\address{Department of Physics, \'{}Transilvania\'{} University, Bd. Eroilor 29,\\
R-2200, Brasov, Romania, e-mail: s.dumitru@unitbv.ro}
\maketitle

\begin{abstract}
In spite \smallskip of their popularity the {\bf H}eisenberg's
(``uncertainty'') {\bf R}elations (HR) still generate controversies. The 
{\bf T}raditional {\bf I}nterpretation of HR\ (TIHR) dominate our days
science, although over the years a lot of its defects were signaled. These
facts justify a reinvestigation of the questions connected with the
interpretation / significance of HR. Here it is developped such a
reinvestigation starting with a revaluation of the main elements of TIHR. So
one finds that all the respective elements are troubled by insurmountable
defects. Then it results the indubitable failure of TIHR and the necessity
of its abandonment. Consequently the HR must be deprived of their quality of
crucial physical formulae. Moreover the HR are shown to be nothing but
simple fluctuations formulae with natural analogous in classical
(non-quantum) physics. The description of the maesuring uncertainties
(traditionally associated with HR) is approached from a new informational
perspective. The Planck's constant $\hbar $ (also associated with HR) is
revealed to have a significance of generic indicator for quantum
stochasticity, similarly with the role of Boltzmann's constant k in respect
with the thermal stochasticity. Some other adjacent questions are also
briefly discussed in the end.
\end{abstract}

\tableofcontents

\smallskip Motto: {\it ''uncertainty principle: it has to do with the
uncertainty in predictions rather than the accuary of measurement. I think
in fact that the word ''measurement'' has been so abused in quantum
mechanics that it would be good to avoid it altogether''}

John S. Bell, 1985.

\section{INTRODUCTION}

\label{sec:introduc}

The {\bf H}eisenberg's (or uncertainty) {\bf R}elations (HR) have a large
popularity, being frequently regarded as crucial formulae of physics or
(Martens 1991) even as expression of ''the most important principle of the
twentieth century physics''. Nevertheless today one knows (Bunge 1977) that
HR ''are probably the most controverted formulae in the whole of the
theoretical physics''. The controversies originate in the association of the
(supposed special) characteristics of measurements at atomic scale with HR
respectively with the foundation and interpretation of quantum theory. The
respective association was initiated and especially sophisticated within the 
{\bf T}raditional ( conventional or orthodox) {\bf I}nterpretation of{\bf \
HR} (TIHR). Very often the TIHR\ is amalgamated with the so-called
Copenhagen interpretation of quantum mechanics.

Elements of the alluded association were preserved one way or another in
almost all investigations of HR subsequent to TIHR. It is notable that, in
spite of their number and variety, the mentioned investigations have not yet
solved in essence the controversies connected with TIHR. But, curiously,
today, large classes of publications and scientists seem to omit (or even to
ignore) discussions about the controversies and defects characterizing the
TIHR. So, tacitly, in our days TIHR seems to remain a largely adopted
doctrine which dominates the questions regarding the foundation and
interpretation of quantum theory. For all that (Piron 1982) ''the idea that
there are defects in the foundations of orthodox quantum theory is
unquestionable present in the conscience of many physicists''.

No doubt, first of all, the above quoted idea regards questions connected
with TIHR. Then the respective questions require further studies and
probably new views. We believe that a promising strategy to satisfy such
requirements is to develop an investigation guided by the goals presented
under the following {\bf P}oints ({\bf P}):

${\bf \underline{P-1.1}}$ : From the vague multitude of sophisticated
statements of TIHR to identify its main elements (hypotheses,
arguments/motivations and assertions).$\blacktriangle $

${\bf \underline{P-1.2}}$ : To add together the significant defects of TIHR
located in connection with the above mentioned elements.$\blacktriangle $

${\bf \underline{P-1.3}}$ : To examine the verity of the respective defects
as well as their significance with respect to TIHR.$\blacktriangle $

${\bf \underline{P-1.4}}$ : To see if such an examination defends TIHR or
irrefutably pleads against it.$\blacktriangle $

${\bf \underline{P-1.5}}$ : In the latter case to admit the failure of TIHR
and to abandon it as an incorrect and useless doctrine.$\blacktriangle $

${\bf \underline{P-1.6}}$ : To see if HR are veritable physical formulae.$%
\blacktriangle $

${\bf \underline{P-1.7}}$ : To search for a genuine reinterpretation of HR.$%
\blacktriangle $

${\bf \underline{P-1.8}}$ : To give a (first) evaluation of the direct
consequences of the mentioned reinterpretation.$\blacktriangle $

${\bf \underline{P-1.9}}$ : To note a few remarks on some adjacent questions.%
$\blacktriangle $

\smallskip In this paper we wish to develop an investigation of the HR
problematic in the spirit of the above mentioned points ${\bf P-1.1}${\bf \
--- }${\bf P-1.8}${\bf .} For such a purpose we will appeal to some elements
(ideas and results) from our works published in last two decades (Dumitru
1974 a, 1974 b, 1977, 1980, 1984, 1987, 1988, 1989, 1991, 1993, 1996, 1999;
Dumitru and Verriets 1995). But here we strive to incorporate the respective
elements into a more argued and elaborated approach. Also we try to make our
exposition as self-contained as possible so that the reader should find it
sufficiently meaningful and persuasive without any important appeals to
other texts.

Through the announced investigation we shall find that all the main elements
of TIHR are affected by insurmountable defects. Therefore we shall reveal
the indubitable failure of TIHR and the necessity of its abandonment. Then
it directly follows that in fact HR do not have any significance connected
with the (measuring) uncertainties. That is why in this paper for the
respective relations we do not use the wide-spread denomination of
''uncertainty relations''.

A consequence of the above alluded revelations is the fact that HR must be
deprived of their quality of crucial physical formulae. So we come in
consonance with the guess (Dirac 1963) that: ''uncertainty relations in
their present form will not survive in the physics of future''.

The failure of TIHR leaves open a conceptual space which firstly requires
justified answers to the questions from ${\bf P-1.6}$ and ${\bf P-1.7}$. The
respective answers must be incorporated in a concordant view about the
subjects of the following points:

${\bf \underline{P-1.10}}$ : The genuine description of the measurements.$%
\blacktriangle $

${\bf \underline{P-1.11}}$ : The foundation and the interpretation of the
actually known quantum theory.$\blacktriangle $

The above mentioned subjects were amalgamated by TIHR trough a lot of
assertions/assumptions which now appear as fallacious. That is why we
suggest that an useful view to be built on a natural differentiation of the
respective subjects.

In such a view the actual quantum theory must be considered as regarding
only intrinsic properties of the entities (particles and fields) from the
microworld. The aspects of the respective properties included in the
theoretical version of HR refer to the stochastic characteristics of the
considered entities. But note that stochastic attributes are specific also
in the case some macroscopic physical systems (e.g. thermodynamical ones),
characterized by a class of macroscopic formulae similar with HR. Also the
Planck's constant $\hbar $ (involved in quantum HR) proves itself to be
similar with the Boltzmann's constant k (involved in the mentioned
macroscopic formulae). Both mentioned constants appear as generic indicators
of stochasticity.

In the spirit of the above suggested view the description of the
measurements remains a question which is extrinsic as regards the properties
of the considered physical systems. Also it must be additional and
independent from the actually known branches of theoretical physics
(including the quantum mechanics). The respective branches refer only to the
intrinsic properties of the considered systems. Then the measurements appear
as processes which supply out-coming (received) information/data about the
intrinsic properties of the measured systems. So regarded the measurements
can be described through some mathematical models. In such models the
measuring uncertainties can be described by means of various estimators.

The above announced views about the HR problematic facilitate
reconsiderations and (we think) nontrivial comments about some questions
regarding the foundations of quantum mechanics.

For developing our exposition in the next sections we will quote directly
only a restricted number of references. This because our goal is not to give
on exhaustive review of the literature dealing with TIHR. The readers
interested in such reviews are invited to consult the known monographical
and bibliographical publications (e.g.: Jammer, 1966; De Witt and Graham,
1971; Jammer, 1974; Nilson, 1976; Yanase {\it et. al.} 1978; Primas, 1981;
Ballentine, 1982; Cramer, 1986; Dodonov and Man'ko, 1987; Martens, 1991;
Braginski and Khalili, 1992; Omnes 1992,1994; Bush {\it et. al}., 1996).

\section{THE MAIN ELEMENTS OF T I H R}

In spite of its popularity, in its promoting literature, TIHR is reported
rather as a vague multitude of sophisticated statements but not as a
systematized ensemble of clearly defined main elements (hypotheses,
arguments/motivations and assertions). However, from the respective
publications there can be identified and sorted out such on ensemble which,
in our opinion, can be presented as follows:

On the best authority (Heisenberg, 1977) today it is known that the TIHR
story originates in the search of general answers to the primary questions
mentioned under the following points:

\underline{${\bf P-2.1}$} ${\bf :}$Are all measurements affected by
measuring uncertainties ?$\blacktriangle $

\underline{${\bf P-2.2}$} ${\bf :}$How can the respective uncertainties be
represented quantitatively in a mathematical scheme ?$\blacktriangle
\medskip $

In connection with{\bf \ }${\bf P-2.1}${\bf ,}TIHR adopted the following
hypotheses:

\underline{${\bf P-2.3}$} ${\bf :}$The measuring uncertainties are due to
the perturbations of the measured system as a result of its interactions
with the measuring instrument.$\blacktriangle $

\underline{${\bf P-2.4}$} ${\bf :}$In the case of macroscopic systems the
mentioned perturbations can be made arbitrarily small and, consequently,
always the corresponding uncertainties can be considered as negligible.

\underline{${\bf P-2.5}$} ${\bf :}$In the case of quantum systems
(microparticles of atomic size) the alluded perturbations are essentially
unavoidable and consequently for certain measurements (see below ${\bf P-2.12%
}$) the corresponding uncertainties are non-negligible.$\blacktriangle $

In the shadow of the hypotheses mentioned in ${\bf P-2.4}$ and ${\bf P-2.5}$
the TIHR attention was limited only to the quantum cases. For approaching
such cases with respect to ${\bf P-2.4}$ TIHR restored to the following
motivation resources:

${\bf \underline{P-2.6}:}$Analysis of some thought (gedanken) measuring
experiments.$\blacktriangle $

${\bf \underline{P-2.7}}$ : Appeal to some theoretical formulae from the
existing quantum mechanics.$\blacktriangle $

The two resources were used in undisguised association. So, from the
starting-point, in TIHR the questions regarding the description of the
measurements, respectively the foundation and interpretation of the existing
quantum theory, were explicitly amalgamated..

For accuracy of the discussions in the following we shall use the term {\it %
variable} in order to denote a physical quantity which describes a specific
property/characteristic of a physical system. With adequate delimitations
the respective term will be used in both theoretical and experimental sense.
In the former case it is connected with the theoretical modeling of system.
In the latter case it is related with the data given by measurements about
the system.

In connection with ${\bf P-2.6}$ there was considered (Heisenberg, 1927,
1930) the case of simultaneous measurements of two (canonically) conjugated
quantum variables A and B (such are coordinate q and momentum p or time t
and energy E). The correspondingly Thought Experimental (TE) uncertainties
are $\Delta _{TE}A$ and $\Delta _{TE}B.$ They were found to be
interconnected trough the following A-B formula

\begin{equation}
\Delta _{TE}A\cdot \Delta _{TE}B\approx \hbar  \eqnum{2.1}
\end{equation}
where $\hbar $ is the quantum Planck's constant.

As regards ${\bf P-2.7}$ firstly there was introduced (Heisenberg 1927,
1930) the following q-p theoretical formula:

\begin{equation}
\Delta _\Psi q\cdot \Delta _\Psi p\geq \frac \hbar 2  \eqnum{2.2}
\end{equation}
(with equality only for Gaussian wave function $\Psi ).$ Afterwards, TIHR
partisans replaced Eq. (2.2) by the more general A-B theoretical formula

\begin{equation}
\Delta _\Psi A\cdot \Delta _\Psi B\geq \frac 12\cdot \left| \left\langle
\left[ \widehat{A},\widehat{B}\right] _{-}\right\rangle _\Psi \right| 
\eqnum{2.3}
\end{equation}
Here $\left[ \widehat{A},\widehat{B}\right] _{-}$ denotes the commutator of
the quantum operators $\widehat{A}$ and $\widehat{B}$, with $\left[ \widehat{%
A},\widehat{B}\right] _{-}=\pm i\hbar $ in the case of conjugated variables.
(For further details about the quantum notations, in actual usance version,
see below the Sec. V).

Equations (2.1) - (2.2)/(2.3) were taken by TIHR as motivation supports.
Based on the such supports TIHR partisans promoted a whole doctrine
(vision). The main (essential) elements of the respective doctrine come down
to the following points, grouped in pairs of Assertions (A) and Motivations
(M):

${\bf \underline{P-2.8/A}}$ : The quantities $\Delta _{TE}A$ and $\Delta
_\Psi A$ from Eqs. (2.1) and (2.2)/(2.3) denoted by an unique symbol $\Delta
A$, have an identical significance of measuring {\it uncertainty} for the
quantum variable A.$\blacktriangle $

${\bf \underline{P-2.8/M}}$ : The above mentioned TIHR presumptions about $%
\Delta _{TE}A$ and the (formal) resemblance between Eqs. (2.1) and (2.2).$%
\blacktriangle $

${\bf \underline{P-2.9/A}}$ : Equations (2.1) and (2.2)/(2.3) admit the same
generic interpretation of uncertainty relations for simultaneous
measurements of the variables $A$ and $B$.$\blacktriangle $

$\underline{{\bf P-2.9/M}}$ : The presumed significance for $\Delta _{TE}A$
and $\Delta _{TE}B$ from Eq. (2.1) and the resemblance between Eqs. (2.1)
and (2.12/(2.3).$\blacktriangle $

${\bf \underline{P-2.10/A}}$ : A solitary quantum variable $A$ can be
measured without any uncertainty (with unlimited accuracy).$\blacktriangle $

${\bf \underline{P-2.10/M}}$ : For such a variable, considered independently
from other variables, the Eqs. (2.1) - (2.3) do not impose a lower bound for
the uncertainty $\Delta A$.$\blacktriangle $

${\bf \underline{P-2.11/A}}$ : Two commutable variables $A$ and $B$ can be
measured simultaneously with arbitrarily small (even null) uncertainties $%
\Delta A$ and $\Delta B$.$\blacktriangle $

${\bf \underline{P-2.11/M}}$ : For such variables $\left[ \widehat{A},%
\widehat{B}\right] _{-}=0$ and in Eq. (2.3) the product $\Delta A\cdot
\Delta B$ has not a lower bound.$\blacktriangle $

${\bf \underline{P-2.12/A}}$ : Two non-commutable variables $A$ and $B$ can
be measured simultaneously only with non-null and interdependent
uncertainties $\Delta A$ and $\Delta B$.$\blacktriangle $

${\bf \underline{P-2.12/M}}$ : In such a case $\left[ \widehat{A},\widehat{B}%
\right] _{-}\neq 0$ and in Eq. (2.3) as well as in Eq. (2.1) the product $%
\Delta A\cdot \Delta B$ of the corresponding simultaneous uncertainties has
as a lower bound a non-null quantity.$\blacktriangle $

${\bf \underline{P-2.13/A}}$ : The HR defined by Eqs. (2.1) - (2.3) (and
named uncertainty relations) are typically quantum formulae and they have no
similar in classical (non-quantum) physics.$\blacktriangle $

${\bf \underline{P-2.13/M}}$ : Presence of the quantum (Planck) constant $%
\hbar $ in Eqs. (2.1) - (2.3) and its absence in all known formulae of
classical physics.$\blacktriangle $

The above mentioned points ${\bf P\dashrightarrow 2.8-P-2.13}$ can be
regarded as main elements of TIHR. This because any piece of the variety of
TIHR statements is obtained and advocated by means of some combinations of
the respective elements.

Among the alluded pieces we mention here the ones regarding the mutual
relations of quantum variables. TIHR adopted the idea:

${\bf \underline{P-2.14}}$ : A variable exists /(can be defined) only when
it is measurable with absolute accuracy (without uncertainty).$%
\blacktriangle $

$\frac {}{}$By combining this idea with ${\bf P-2.11}$ and ${\bf P-2.12}$ in
the TIHR literature it is often promoted the statement:

${\bf \underline{P-2.15}}$ : Two quantum variables are compatible
respectively incompatible as their operators are respectively are not
commutable. Consequently a complete description of a quantum system must be
made in terms of a set of mutually compatible variables.${\bf \blacktriangle 
}$

In the same literature one finds also the opinion that:

${\bf \underline{P-2.16}}$ : Two incompatible variables (especially the
canonically conjugated ones) are complementary (i.e. mutually exclusive) -
similarly as in the complementarity relation for the corpuscular and wave
characteristics of microparticles of atomic size.${\bf \blacktriangle }$

\section{A FEW REMARKS ON T I H R HISTORY}

TIHR was initiated by Heisenberg but later on it was developed and
especially promoted by the Copenhagen School guided by N. Bohr. In a first
stage TIHR had a relatively modest motivation, based only on the Eqs. (2.1)
and (2.2). However it was largely accepted in scientific and academic
communities, partly due to the authority of its promoters. So, the
establishing of TIHR as a doctrine started.

In a second stage TIHR partisans introduced a multitude of
thought-experimental or theoretical formulae which resemble more or less the
Eqs. (2.1)-(2.3). In spite of their (conceptual and/or mathematical)
diversity the respective formulae were declared as {\it uncertainty relations%
} and their existence and interpretation were regarded as supports for an
extended motivation of TIHR. So, for its partisans, TIHR was viewed as a
well established and irrefutable doctrine. Such a view was widely promoted
in leading publications (especially in textbooks).

In the meantime the alluded view was confronted with the notification of
some defects of TIHR. But, as a rule, the respective notifications appeared
disparately, sometimes in marginal publications and from non-leading
authors. So the mentioned defects were not presented as a systematized
ensemble and TIHR was criticized on certain points but not in its totality.
An appreciation viewing somehow the alluded totality was noted altogether
solitarily (Primas 1981). Referring to the post-Copenhagen interpretation of
quantum mechanics it says: ''Heisenberg\'{}s uncertainty relations are no
longer at the heart of the interpretation but are simple consequences of the
basic mathematical formalism''. Here one should remark that, as we know,
such an appreciation has never been used in order to elucidate the
shortcomings of the TIHR doctrine. Moreover it seems that, even in the
our-days publications regarding the interpretation of quantum mechanics, the
respective appreciation is not taken properly into account.

In the presented circumstances the TIHR partisans ignored or even denied the
alluded defects. Such an attitude was sustained mainly by putting forward
thought experiments and/or the authority of the mentioned partisans. But
note that, in this way, for most of the cases, the notifications of TIHR
defects were not really counteracted and the corresponding controversies
were not elucidated. For all that TIHR survived over the decades with the
appearance of an uncontroversial doctrine and became a veritable myth.
Undoubted signs of the respective myth are present even today in
publications (especially in textbooks) and in the thinking of peoples
(particularly in academic communities).

Here, it is interesting to observe Heisenberg\'{}s own attitude towards TIHR
story. It is surprising to see in the afferent literature that, although he
was the initiator of TIHR doctrine, Heisenberg was not involved in the
subsequent history of the respective doctrine. So he did not develop
mathematical generalizations or interpreting extensions/sophistications of
the Eqs. (2.1)-(2.3). Also he did not participate in the controversies
regarding the TIHR defects. Probably that was the reason why in one of his
last publications on HR (Heisenberg, 1977) he did not refer to such
developments and controversies but reminded only his thoughts connected with
the beginning of the TIHR history. Can the alluded attitude be regarded as
an evidence of the supposition that in fact Heisenberg was conscious of the
insurmountable defects of TIHR ? A pertinent answer to such a question is
expected to be (eventually) known by the publication of all volumes of a
planned monography (Mehra and Rechenberg, 1982) due, in part, to one of
Heisenberg\'{}s last collaborators.

With the Heisenberg case one discloses a particularity in the attitude of
many scientists who promoted TIHR. As individuals, each of the respective
scientists did not regard the TIHR as a whole doctrine but argued for only a
few of its elements and ignored almost all of the defects. Often their
considerations were amalgamated with ideas which do not pertain strictly to
TIHR. That is why, probably, by the term {\it TIHR-partisans} it is more
adequate to understand a fuzzy class of people rather than a rigorously
delimited group of scientists.

Now looking back over the time, we believe that the verity and true
significance of TIHR defects still remain open questions which require to be
elucidated.. Such a requirement implies the necessity of an argued and
complete revaluation of TIHR. Then, there directly appears the need for a
search of a genuine reinterpretation of HR. The alluded beliefs will guide
our investigations in the following sections.

\section{STARTING CONSIDERATIONS OF T I H R}

TIHR introduced its main elements presented in ${\bf P-2.8}$ {\bf --- }${\bf %
P-2.13}$ by appealing to some starting considerations about the Eqs.
(2.1)-(2.3). The appeals viewed the scientific achievements from the first
years of quantum mechanics.. Here, for a correct (re)evaluation of TIHR, it
is the place to remember briefly the respective considerations.

Firstly it must be noted that the Eqs. (2.1) were introduced by using the
wave characteristics of quantum microparticles. Consequently the quantum
measurements were regarded by similitude with the optical ones. But in the
mentioned years the performances of the optical measurements were restricted
by the classical limitative criteria of resolution (due to Abbe and
Rayleigh). Then TIHR promoted as starting consideration the following point:

${\bf \underline{P-4.1}}$: The estimation of performances respectively of
uncertainties for the quantum measurements must be done by using the alluded
limitative criteria, transposed in quantum framework through de Broglie
formula $\lambda =h/p$ ($\lambda $ = wave lenght).$\blacktriangle $

By means of this consideration TIHR partisans obtained some relations
similar with Eqs. (2.1), for all the thought-experiments promoted by them.

Referring to the Eqs. (2.2)-(2.3) the starting considerations promoted by
TIHR can be resumed as follows. The state of a quantum microparticle is
described by the wave function $\Psi =\Psi (q)$ regarded as a vector in a
Hilbert space (q denotes the set of specific orbital variables). In the
respective vectorial space the scalar product ($\Psi _a,\Psi _b)$ of two
functions $\Psi _a$ and $\Psi _b$ is given by

\begin{equation}
\left( \Psi _a,\Psi _b\right) =%
\displaystyle \int %
\limits_{\Omega _q}\Psi _a^{*}\Psi _ad\Omega _q  \eqnum{4.1}
\end{equation}

where $\Psi _a^{*}$=the complex conjugate of $\Psi _a$, whereas $\Omega _q$
and $d\Omega _q$ denote the accessible respectively infinitesimal domains in
q-space. $A$ quantum variable $A$ is described by the operator $\widehat{A}$
and its expected (mean) value $\left\langle A\right\rangle _\Psi $ is
defined by

\begin{equation}
\left\langle A\right\rangle _\Psi =\left( \Psi ,\widehat{A}\Psi \right) 
\eqnum{4.2}
\end{equation}
The quantity $\Delta _\Psi A$ from Eqs. (2.2)-(2.3) is defined as follows:

\[
D_\Psi A=\left( \delta _\Psi \widehat{A}\Psi ,\delta _\Psi \widehat{A}\Psi
\right) \,\,\,\,\,,\,\,\,\,\,\,\delta _\Psi \widehat{A}=\widehat{A}%
-\left\langle A\right\rangle _\Psi 
\]
\begin{equation}
\Delta _\Psi A=\sqrt{D_\Psi A}  \eqnum{4.3}
\end{equation}
Then for two variables $A$ and $B$ the following evident relation was
appealed

\begin{equation}
\left( \left( \alpha \delta _\Psi \widehat{A}-i\delta _\Psi \widehat{B}%
\right) \Psi ,\left( \alpha \delta _\Psi \widehat{A}-i\delta _\Psi \widehat{B%
}\right) \Psi \right) \geq 0  \eqnum{4.4}
\end{equation}
with $\alpha $ an arbitrary and real parameter. In the TIHR literature this
relation is transcribed into the formula

\begin{equation}
\left( \Psi ,\left( \alpha \delta _\Psi A+i\delta _\Psi \widehat{B}\right)
\left( \alpha \delta _\Psi \widehat{A}-i\delta _\Psi \widehat{B}\right) \Psi
\right) \geq 0  \eqnum{4.5}
\end{equation}
which is equivalent with the relation

\begin{equation}
\alpha ^2\left( \Delta _\Psi A\right) ^2-\alpha \left\langle i\left[ 
\widehat{A},\widehat{B}\right] _{-}\right\rangle _\Psi +\left( \Delta _\Psi
B\right) ^2\geq 0  \eqnum{4.6}
\end{equation}
where $\left[ \widehat{A},\widehat{B}\right] _{-}=\widehat{A}\widehat{B}-%
\widehat{B}\widehat{A}$ is the commutator of the operators $\widehat{A}$ and 
$\widehat{B}.$ As Eq. (4.6) is satisfied for any value of $\alpha $ it
directly results.

\begin{equation}
\left( \Delta _\Psi A\right) ^2\cdot \left( \Delta _\Psi B\right) ^2\geq 
\frac 14\left\langle i\left[ \widehat{A},\widehat{B}\right]
_{-}\right\rangle _\Psi ^2  \eqnum{4.7}
\end{equation}
or

\begin{equation}
\Delta _\Psi A\cdot \Delta _\Psi B\geq \frac 12\left| \left\langle \left[ 
\widehat{A},\widehat{B}\right] _{-}\right\rangle _\Psi \right|  \eqnum{4.8}
\end{equation}
This latter formula is just the Eq. (2.3).

Now we point out here the following notable facts. The above mentioned
starting considerations were founded on old scientific achievements. For all
that, they are preserved and promoted in an unchanged form even in
today\'{}s literature (especially in textbooks).

\section{ULTERIOR SCIENTIFIC\ ACHIEVEMENTS}

For an up-to-date re-evaluation of TIHR the above mentioned starting
considerations must be supplemented with elements regarding some ulterior
scientific achievements. The respective elements were reported in the
decades after the debut of quantum mechanics but, surprisingly, even today
they seem to have little popularity. Our alluded supplement regards the
following things.

Firstly let us refer to the thought-experimental Eqs. (2.1) and,
correspondingly, to the limitative criteria involved in the starting
consideration mentioned in ${\bf P-4.1}${\bf .} In the last decades, in
optical measurements, some super-resolution techniques have been achieved
(Roychoudhuri, 1978; Scheer {\it et. al.}, 1989; Croca {\it et al}., 1996).
The performances of the respective techniques overstep the alluded
limitative criteria. Then it seems to be possible that instead of ${\bf P-4.1%
}$ one should operate with the following up-to-date consideration:

$\underline{{\bf P-5.1}}$ : The accuracies and uncertainties of the quantum
measurements can be estimated by transposition in adequate terms (by means
of de Broglie formula $\lambda =h/p$)of the mentioned super-resolution
performances.$\blacktriangle $

Based on this consideration, it is easy to imagine some{\bf \ S}uper-{\bf R}%
esolution {\bf T}hought {\bf E}xperiments (SRTE). Then for the measurement
of two variables $A$ and $B$ the corresponding uncertainties are $\Delta
_{SRTE}A$ and $\Delta _{SRTE}B.$ By rationating similarly as in the case of
Eq. (2.1) one finds 
\begin{equation}
\Delta _{SRTE}A\cdot \Delta _{SRTE}B<\hbar  \eqnum{5.1}
\end{equation}
This $SRTE$ - relation must be taken into account for un up-to-date
re-evaluation of TIHR.

Now let us refer to some ulterior achievements connected with the
theoretical Eqs. (2.2)-(2.3) or (4.8). The respective achievements regard
mathematical generalizations of the Eq. (4.8). Note that there is known
(Dodonov and Man\'{}ko, 1987; Dumitru, 1988) a large variety of such
generalized relations. But here we shall consider only a few of the
respective relations which are of direct significance for the questions
approached in this paper.

Then we consider a quantum microparticle for which the orbital state and
variables are described by the wave function $\Psi $ respectively by the
operators $\widehat{A}_k(k=1,2,.....n).$ With the same significance of
notations as in Eqs. (4.1)-(4.4), we can define the correlations:

\[
C_\Psi \left( A_jA_k\right) =\left( \delta _\Psi \widehat{A}_j\Psi ,\delta
_\Psi \widehat{A}_k\Psi \right) 
\]
\begin{equation}
D_\Psi A_j=C_\Psi \left( A_jA_j\right) \,\,\,\,,\,\,\,\Delta _\Psi A_j=\sqrt{%
D_\Psi A_j}  \eqnum{5.2}
\end{equation}
If $\alpha _k\left( k=1,2,.....,n\right) $ are a set of arbitrary and
complex parameters we can write the evident relation

\begin{equation}
\left( \sum_{k=1}^n\alpha _k\delta _\Psi \widehat{A}_k\Psi
,\sum_{l=1}^n\alpha _l\delta _\Psi \widehat{A}_l\Psi \right) \geq 0 
\eqnum{5.3}
\end{equation}
which can be transcribed directly as: 
\begin{equation}
\sum_k\sum_l\alpha _k^{*}\alpha _l\left( \delta _\Psi \widehat{A}_k\Psi
,\delta _\Psi \widehat{A}_l\Psi \right) \geq 0  \eqnum{5.4}
\end{equation}
The quantities $\left( \delta _\Psi \widehat{A}_k\Psi ,\delta _\Psi \widehat{%
A}_l\Psi \right) (k;l=1,2,...,n)$ represents the {\it correlation matrix} of
the set of variables $A_k$. It is obvious that

\begin{equation}
\left( \delta _\Psi \widehat{A}_k\Psi ,\delta _\Psi \widehat{A}_l\Psi
\right) ^{*}=\left( \delta _\Psi \widehat{A}_l\Psi ,\delta _\Psi \widehat{A}%
_k\Psi \right)  \eqnum{5.5}
\end{equation}
i.e. the correlation matrix is Hermitian. Equation (5.4) shows that the
respective matrix is also non-negative definite. Then from the matrix
algebra (see Korn and Korn, 1968) it results 
\begin{equation}
\det \left[ \left( \delta _\Psi \widehat{A}_k\Psi ,\delta _\Psi \widehat{A}%
_l\Psi \right) \right] \geq 0  \eqnum{5.6 $_{CR}$}
\end{equation}
where $\det \left[ a_{kl}\right] $ denotes the determinant with the elements 
$a_{kl}.$ Here, and in the following notations, the index CR added to the
number of a formula shows the belonging of the respective formula to a
general family of similar {\it correlation relations} (CR).

For two operators $\widehat{A}_1=\widehat{A}$ and $\widehat{A}_2=\widehat{B}$
from Eq. (5.6) one obtains

\begin{equation}
\left( \delta _\Psi \widehat{A}\Psi ,\delta _\Psi \widehat{A}\Psi \right)
\cdot \left( \delta _\Psi \widehat{B}\Psi ,\delta _\Psi \widehat{B}\Psi
\right) \geq \left| \left( \delta _\Psi \widehat{A}\Psi ,\delta _\Psi 
\widehat{B}\Psi \right) \right| ^2  \eqnum{5.7 $_{CR}$}
\end{equation}
If the two operators satisfy the conditions

\begin{equation}
\left( \widehat{A}_k\Psi ,\widehat{A}_l\Psi \right) =\left( \Psi ,\widehat{A}%
_k\widehat{A}_l\Psi \right) \left( k=1,2;l=1,2)\right)  \eqnum{5.8}
\end{equation}
equation (5.6) gives directly

\begin{equation}
\Delta _\Psi A\cdot \Delta _\Psi B\geq \left| \left\langle \delta _\Psi 
\widehat{A}\cdot \delta _\Psi \widehat{B}\right\rangle _\Psi \right| 
\eqnum{5.9 $_{CR}$}
\end{equation}
When Eq. (5.8) is satisfied we have also

\begin{equation}
\left\langle \delta _\Psi \widehat{A}\delta _\Psi \widehat{B}\right\rangle
_\Psi =\frac 12\left\langle \left[ \delta _\Psi \widehat{A},\delta _\Psi 
\widehat{B}\right] _{+}\right\rangle _\Psi -\frac i2\left\langle i\left[ 
\widehat{A},\widehat{B}\right] \right\rangle _\Psi  \eqnum{5.10}
\end{equation}
where $\left[ \widehat{A},\widehat{B}\right] _{\pm }=\widehat{A}\widehat{B}%
\pm \widehat{B}\widehat{A}$ (i.e. the anticomutator respectively commutator
of $\widehat{A}$ and $\widehat{B}$) while $\left\langle \left[ \delta _\Psi 
\widehat{A},\delta _\Psi \widehat{B}\right] _{+}\right\rangle _\Psi $ and $%
\left\langle i\left[ \widehat{A},\widehat{B}\right] _{-}\right\rangle _\Psi $
are real quantities. Then Eq. (5.9) can be transcribed as

\begin{equation}
\Delta _\Psi A\cdot \Delta _\Psi B\geq \sqrt{\frac 14\left\langle \left[
\delta _\Psi \widehat{A},\delta _\Psi \widehat{B}\right] _{+}\right\rangle
_\Psi ^2+\frac 14\left\langle i\left[ \widehat{A},\widehat{B}\right]
_{-}\right\rangle _\Psi ^2}  \eqnum{5.11 $_{CR}$}
\end{equation}
This formula implies the following two less restrictive relations

\begin{equation}
\Delta _\Psi A\cdot \Delta _\Psi B\geq \frac 12\left\langle \left[ \delta
_\Psi \widehat{A},\delta _\Psi \widehat{B}\right] _{+}\right\rangle _\Psi 
\eqnum{5.12 $_{CR}$}
\end{equation}
and 
\begin{equation}
\Delta _\Psi A\cdot \Delta _\Psi B\geq \frac 12\left| \left\langle \left[ 
\widehat{A},\widehat{B}\right] _{-}\right\rangle _\Psi \right| 
\eqnum{5.13
$_{CR}$}
\end{equation}
One can see that the latter relation is exactly the theoretical version from
Eqs.(2.3)/(4.8) of HR.

Note now that there are situations when the Eq. (5.8) is satisfied for $k=l$
but not for $k\neq l$. In such situations from Eq. (5.7) instead of Eq.
(5.9) one obtains the relation

\begin{equation}
\Delta _\Psi A\cdot \Delta _\Psi B\geq \left| \left( \delta _\Psi \widehat{A}%
\Psi ,\delta _\Psi \widehat{B}\Psi \right) \right|  \eqnum{5.14 $_{CR}$}
\end{equation}

From the above presented considerations it results that the true
generalizations of the theoretical HR given by Eqs. (2.2/(2.3)/(4.8)/(5.13)
are exactly the Eqs. (5.7) and (5.6).

The above discussed relations refer to the orbital variables of a quantum
microparticle. But such a microparticle has also spin variables
characterized by similar relations. So if for an electron the spin state and
the spin variables are described by the spinor $\chi $ (spin wave function)
respectively by the matrices (operators) $\widehat{A}_j$ the alluded
relations can be introduced as follows. With the usual notations (see
Bransden and Joachain, 1994) the expected values are $\left\langle
A_j\right\rangle _\chi =\chi ^{+}A_j\chi $ while the correlations $C_\chi
\left( A_jA_e\right) ,$ dispersions $D_\chi A_j$, and standard deviation $%
\Delta _\chi A_j$ are given by

\[
C_\chi \left( A_jA_l\right) =\left( \delta _\chi \widehat{A}_j\chi \right)
^{+}\cdot \left( \delta _\chi \widehat{A}_l\chi \right) \text{,\thinspace
\thinspace \thinspace \thinspace \thinspace \thinspace \thinspace \thinspace
\thinspace \thinspace \thinspace \thinspace \thinspace \thinspace \thinspace
\thinspace \thinspace \thinspace \thinspace \thinspace \thinspace \thinspace
\thinspace \thinspace \thinspace \thinspace \thinspace \thinspace }\delta
_\chi \widehat{A}_j=\widehat{A}_j-\left\langle A_j\right\rangle _\chi 
\]
\begin{equation}
D_\chi A_j=C_\chi \left( A_jA_j\right) \text{,\thinspace \thinspace
\thinspace \thinspace \thinspace \thinspace \thinspace \thinspace \thinspace
\thinspace \thinspace \thinspace \thinspace \thinspace \thinspace \thinspace
\thinspace \thinspace \thinspace \thinspace \thinspace \thinspace \thinspace 
}\Delta _\chi A_j=\sqrt{D_\chi A_j}  \eqnum{5.15}
\end{equation}
Similarly to the orbital Eqs. (5.3)-(5.6) it is easy to see that the
spin-correlations $C_\chi \left( A_jA_l\right) $satisfy the relation 
\begin{equation}
\det \left[ C_\chi \left( A_jA_l\right) \right] \geq 0  \eqnum{5.16 $_{CR}$}
\end{equation}
For two variables $A_1=A$ and $A_2=B$, which satisfy conditions similar to
(5.8), from (5.16) one obtains

\begin{equation}
\Delta _\chi A\cdot \Delta _\chi B\geq \left| \left\langle \delta _\chi 
\widehat{A}\delta _\chi \widehat{B}\right\rangle _\chi \right| 
\eqnum{5.17$_{CR}$}
\end{equation}
\begin{equation}
\Delta _\chi A\cdot \Delta _\chi B\geq \frac 12\left| \left\langle \left[ 
\widehat{A},\widehat{B}\right] _{-}\right\rangle _\chi \right| 
\eqnum{5.18
$_{CR}$}
\end{equation}
Equations (5.18) and (5.16) are nothing but spin similars of orbital HR and
of their generalizations given by Eqs. (5.13) respectively (5.6).

Theoretical versions of HR (as well as their generalizations) imply, for the
variables of the quantum microparticles, a lot of probabilistic parameters
such as: expected/mean values, dispersions, standard deviations and
correlations. This means that the respective variables have stochastic (or
random) characteristics. Then there follows directly the question: are there
similars of HR for other physical systems, different from quantum
microparticles, which have also variables with stochastic characteristics ?
The answer to the mentioned questions is affirmative and it regards
macroscopic systems studied in both classical and quantum statistical
physics. Here we shall illustrate the respective answer by taking over some
ideas from our earlier works (Dumitru, 1974a, 1977, 1988, 1993).

Firstly let us refer to a macroscopic system, consisting of a large number
of microparticles, considered in a thermodynamic equilibrium state. In the
framework of classical (nonquantum) statistical physics such a system can be
approached in terms of: (a) phenomenological (quasithermodynamic)
fluctuations theory respectively (b) classical statistical mechanics. In the
mentioned approaches the state of the system is described (Landau and
Lifchitz, 1984; Zubarev, 1971; Ruppeiner, 1995) by the distribution function 
$w=w(x)$. The variable $x$ denotes in the two cases: (a) the set of
independent macroscopic variables of the system as a whole, respectively (b)
the phase space coordinates of the microparticles constituting the system.
In both cases a specific variable $A$ characterizing the system is a real
stochastic (random) quantity with a continuous spectrum of values which
depends on $x$, i.e. $A=A(x)$. The mean (or expected) value of $A(x)$ is
given by

\begin{equation}
\left\langle A\right\rangle _w=\int_{\Omega _X}A\left( x\right) w\left(
x\right) d\Omega _x  \eqnum{5.19}
\end{equation}
where $\Omega _x$ and $d\Omega _x$ denote the accessible respectively
infiniesimal domains in the $x$-space. Then, in the case of such a
macroscopic system, for a set $A_j(j=1,2,...,n)$ of specific variables, the 
{\it thermal fluctuations} are described by the correlations $C_w\left(
A_jA_e\right) $, dispersions $D_wA_j$ and standard deviations $\Delta _wA_j$
given by

\[
C_w\left( A_jA_l\right) =\left\langle \delta _wA_j\delta _wA_l\right\rangle
_w,\,\,\,\,\,\,\,\,\,\,\,\,\,\delta _wA_j=A_j-\left\langle A_j\right\rangle
_w 
\]
\begin{equation}
D_wA_j=C_w\left( A_jA_j\right) ,\,\,\,\,\,\,\,\,\,\,\,\,\,\,\,\Delta _wA_j=%
\sqrt{D_wA_j}  \eqnum{5.20 }
\end{equation}
By relations similar to Eqs.(5.3)-(5.5) it is easy to see that the
correlations $\left\langle \delta _wA_j\delta _wA_l\right\rangle $ are the
elements of a non-negative real matrix. Then similarly with Eq. (5.6) one
can write

\begin{equation}
\det \left[ \left\langle \delta _wA_j\delta _wA_l\right\rangle _w\right]
\geq 0  \eqnum{5.21$_{CR}$}
\end{equation}
Particularly for two variables $A_1=A$ and $A_2=B$ one obtains

\begin{equation}
D_wA\cdot D_wB\geq \left\langle \delta _wA\cdot \delta _wB\right\rangle _w^2
\eqnum{5.22 $_{CR}$}
\end{equation}
\begin{equation}
\Delta _wA\cdot \Delta _wB\geq \left| \left\langle \delta _wA\cdot \delta
_wB\right\rangle _w\right|  \eqnum{5.23 $_{CR}$}
\end{equation}
Equations (5.20)-(5.22) can be called {\it thermal correlation relations. }%
Some examples of such relations are given below in Sec. VI. K. (see also
Dumitru, 1974a, 1988, 1993).

One observes that Eqs. (5.23) and (5.21) are the macroscopic similars of
microscopic HR and of their generalizations defined by Eqs.
(2.3)/(4.8)/(5.13) respectively (5.6). Here it must be noted that there are
also other macroscopic similars of HR, namely relations from the framework
of quantum statistical mechanics. Such relations can be obtained as follows:
The state respectively the specific variables of a macroscopic system in the
mentioned framework are described by the statistical operator (density
matrix) $\widehat{\rho }$ respectively by the operators $\widehat{A}%
_j\,\,(j=1,2,...,n).$ With the expected values defined as $\left\langle
A_j\right\rangle _\rho =T_r\left( \widehat{A}_j\widehat{\rho }\right) $ the
macroscopic correlations $C_\rho \left( A_jA_e\right) ,$ dispersions $D_\rho
A_j$ and standard deviations $\Delta _\rho A_j$ are given by

\[
C_\rho \left( A_jA_e\right) =\left\langle \delta _\rho \widehat{A}_j\cdot
\delta _\rho \widehat{A}_l\right\rangle _\rho ,\,\,\,\,\,\,\,\delta _\rho 
\widehat{A}_j=\widehat{A}_j-\left\langle A_j\right\rangle _{\rho
\,}\,\,\,\,\,\, 
\]
\begin{equation}
D_\rho A_j=C_\rho \left( A_jA_j\right) ,\,\,\,\,\,\ \Delta _\rho A_j=\sqrt{%
D_\rho A_j}  \eqnum{5.24}
\end{equation}
In sufficiently general circumstances (among them the most important being
some conditions similar with Eq.(5.8)) the quantities from Eqs.(5.24)
satisfy the relations

\begin{equation}
\det \left[ \left\langle \delta _\rho \widehat{A}_j\delta _\rho \widehat{A_l}%
\right\rangle _\rho \right] \geq 0  \eqnum{5.25}
\end{equation}
\begin{equation}
D_\rho A\cdot D_\rho B\geq \left| \left\langle \delta _\rho A\cdot \delta
_\rho B\right\rangle _\rho \right| ^2  \eqnum{5.26 $_{CR}$}
\end{equation}
\begin{equation}
\Delta _\rho A\cdot \Delta _\rho B\geq \left| \left\langle \delta _\rho
A\cdot \delta _\rho B\right\rangle _\rho \right|  \eqnum{5.27$_{CR}$}
\end{equation}
with $\widehat{A}=\widehat{A}_1$, $\widehat{B}=\widehat{A}_2$. From
Eq.(5.27) one obtains also the following truncated (less restrictive)
relations

\begin{equation}
\Delta _\rho A\cdot \Delta _\rho B\geq \frac 12\left\langle \left[ \delta
_\rho \widehat{A},\delta _\rho \widehat{B}\right] _{+}\right\rangle _\rho 
\eqnum{5.28$_{CR}$}
\end{equation}
\begin{equation}
\Delta _\rho A\cdot \Delta _\rho B\geq \frac 12\left| \left\langle \left[ 
\widehat{A},\widehat{B}\right] _{-}\right\rangle _\rho \right| 
\eqnum{5.29
$_{CR}$}
\end{equation}
The latter relation is exactly a macroscopic similar of HR defined by Eqs.
(2.3)/(4.8)/(5.13).

The above discussed relations are unitemporal in the sense that the implied
probabilistic parameters (correlations, dispersions, standard deviations) of
the stochastic variables $A_j$ are considered for the same moment of time.
But it easy to see that similar relations can be written if the mentioned
parameters are taken into account for different time moments. So, if the
orbital quantum state of a microparticle is described by the time t
dependent wave function $\Psi \left( q,t\right) $, instead of unitemporal
Eq. (5.14) one can write the following bitemporal relation:

\begin{equation}
\Delta _{\Psi _1}A\cdot \Delta _{\Psi _2}B\geq \left| \left( \delta _{\Psi
_1}\widehat{A}\Psi _1,\delta _{\Psi _2}\widehat{B}\Psi _2\right) \right| 
\eqnum{5.30}
\end{equation}
where $\Psi _1=\Psi \left( q,t_1\right) $ and $\Psi _2=\Psi \left(
q,t_2\right) $ with $t_1\neq t_2$.

Another well-known way of introducing the theoretical HR for orbital
variables is based on Fourier analysis as follows. Let be $f(x)$ a
continuous and quadratically integrable function in the range $x\in \left(
-\infty ,\infty \right) $. Then its Fourier transforms $\widetilde{f}\left(
k\right) $ is defined by: 
\begin{equation}
\widetilde{f}\left( k\right) =\frac 1{\sqrt{2\pi }}\int_{-\infty }^\infty
f\left( x\right) e^{-ikx}dx  \eqnum{5.31}
\end{equation}
If $\left| f\left( x\right) \right| ^2$ and $\left| \widetilde{f}\left(
k\right) \right| ^2$ are normalized to unity by using the Parseval formula
one can write 
\begin{equation}
\int_{-\infty }^\infty \left| f\left( x\right) \right| ^2dx=\int_{-\infty
}^\infty \left| f\left( k\right) \right| ^2dk=1  \eqnum{5.32}
\end{equation}
Then $\left| f\left( x\right) \right| ^2$ and $\left| \widetilde{f}\left(
k\right) \right| ^2$ can be interpreted as probability densities for the
variables $x$ respectively $k$. Consequently the mean (expected) values of
functions like $A(x)$ or $B(k)$ are given by 
\begin{equation}
\left\langle A\left( x\right) \right\rangle =\int_{-\infty }^\infty A\left(
x\right) \left| f\left( x\right) \right| ^2dx  \eqnum{5.33}
\end{equation}
\begin{equation}
\left\langle B\left( k\right) \right\rangle =\int_{-\infty }^\infty B\left(
k\right) \left| \widetilde{f}\left( k\right) \right| ^2dk  \eqnum{5.34}
\end{equation}
With the above mentioned elements one can demonstrate (De Bruijn, 1967) the
following relation 
\begin{equation}
\left\langle \left( x-a\right) ^2\right\rangle \cdot \left\langle \left(
k-b\right) ^2\right\rangle \geq \frac 14  \eqnum{5.35}
\end{equation}
with {\it a} and {\it b} as two arbitrary constants.

One observes that from Eq. (5.35) it results directly the HR defined by Eq.
(2.2), for the Cartesian coordinate $x$ and momentum p. For such a result
one must take $f\left( x\right) $ as the wave function $\Psi \left( x\right) 
$ and respectively $a=\left\langle x\right\rangle $, $b=\left\langle
k\right\rangle $ and $k=p/\hbar $.

If $f\left( x\right) $ is periodic in $x$, with the period $\Lambda $, or is
defined in the range $x\in \left[ x_0,\Lambda \right] $ with the same values
on the boundaries, it satisfies the relation 
\begin{equation}
f\left( x_0\right) =f\left( x_0+\Lambda \right)  \eqnum{5.36}
\end{equation}
Then instead of $\widetilde{f}\left( k\right) $ from Eq. (5.31) we have the
Fourier coefficients 
\begin{equation}
\widetilde{f}_n=\frac 1{\sqrt{\Lambda }}\int_{x_0}^{x_0+\Lambda }f\left(
x\right) e^{-ik_nx}dx  \eqnum{5.37}
\end{equation}
with $k_n=n\cdot 2\pi /\Lambda $ and $n=0,\pm 1,\pm 2,......$ Similarly with
Eq. (5.33) we can take: 
\begin{equation}
\int_{x_0}^{x_0+\Lambda }\left| f\left( x\right) \right|
^2dx=\sum_{n=-\infty }^\infty \left| \widetilde{f}_n\right| ^2=1 
\eqnum{5.38}
\end{equation}
This means that $\left| f\left( x\right) \right| ^2$ can be interpreted as
probability density for the continuous variable $x$ while $\left| \widetilde{%
f}_n\right| ^2$ signify the probabilities associated with the discrete
variable $k_n.$ In such a case instead of Eqs. (5.33) and (5.34) one must
write 
\begin{equation}
\left\langle A\left( x\right) \right\rangle =\int_{x_0}^{x_0+\Lambda
}A\left( x\right) \left| f\left( x\right) \right| ^2dx  \eqnum{5.39}
\end{equation}
\begin{equation}
\left\langle B\left( k\right) \right\rangle =\sum_{n=-\infty }^\infty
B\left( k_n\right) \left| \widetilde{f}_n\right| ^2,\,\,\,\,\,\,\,\,k_n=n%
\frac{2\pi }\Lambda  \eqnum{5.40}
\end{equation}
and instead of Eq. (5.35) one obtains 
\begin{equation}
\left\langle \left( x-a\right) ^2\right\rangle \cdot \left\langle \left(
k-b\right) ^2\right\rangle =\frac 14\left| \left( \Lambda f\left( x_0\right)
-1\right) \right| ^2  \eqnum{5.41$_{CR}$}
\end{equation}
This latter formula is applicable in some cases for the variables azimuthal
angle $\varphi $ and angular momentum $L_z$ (see below the Sec. VI.F). In
such cases $f\left( x\right) $ is the periodic wave function $\Psi \left(
\varphi \right) $and respectively $a=\left\langle x\right\rangle
,\,b=\left\langle k\right\rangle $ with $x=\varphi ,\,k=L_z/\hbar $ and $%
\Lambda =2\pi .$

We end this section with a notification regarding the relations expressed by
Eqs.: (5.6), (5.9), (5.11)-(5.14), (5.16)-(5.18), (5.21)-(5.23),
(5.25)-(5.29), (5.30), (5.35) and (5.41). From a mathematical viewpoint all
the respective relations refer to variables with stochastic characteristics.
Also, by their mathematical significances, they belong to the same family of
similar formulae which can be called {\it correlation relations} (CR). That
is why we added the index CR to the numbers of all the respective relations.

\section{DEFECTS OF\ T I H R}

With the above mentioned facts now we can proceed to present the defects of
TIHR. Note that the respective defects appear not as a systematized ensemble
but rather as a dispersed set of (relatively) distinct cases. That is why
our approach aims not at a precisely motivated order of exposition. We
mostly wish to show that taken together the set of the alluded defects
irrefutably incriminate all the main elements of TIHR reviewed in Sec. II.
Then our presentation includes the defects revealed in the following
sub-sections:

\subsection{Groundlessness of the term ''uncertainty''}

Trough the assertion ${\bf P-2.8/A}$ of TIHR the thought respectively
theoretical quantities $\Delta _{TE}A$ and $\Delta _\Psi A$ from HR are
termed measuring uncertainties. But the respective term appears as
groundless if it is regarded comparatively with a lot of facts which we
present here.

Firstly note that a minute examination of all thought experiments referred
in connection with Eq. (2.1) does not justify the mentioned term for one of
the implied quantities $\Delta _{TE}A$ and $\Delta _{TE}B.$ So $\Delta
_{TE}p $ (in the coordinate q -momentum p case) and $\Delta _{TE}E$ (in the
time t - energy E case) represent the jumps of the respective variable from
an initial value (before the measurements) to a final value (after the
measurements). Then it results that $\Delta _{TE}p$ and $\Delta _{TE}E$ can
not be regarded as uncertainties (i.e. measuring parameters) with respect to
the measured state which is the initial one. This because (Albertson, 1963):
''it seems essential to the notion of a measurements that it answer a
question about the given situation existing before measurement. Whether the
measurement leaves the measured system unchanged or brings about a new and
different state of that system is a second and independent question''.

The remaining quantities $\Delta _{TE}q$ and $\Delta _{TE}t$ from Eq. (2.1)
are also in the situation of infringing the term ''measuring uncertainties''
in the sense attributed by TIHR. The respective situation is generated by
the same facts which will be presented below in the Sec. VI.B.

As regards the theoretical quantity $\Delta _\Psi A$ the following
observations must be taken into account. $\Delta _\Psi A$ depends only on
the theoretical model (wave function $\Psi )$ of the considered
microparticle but not on the characteristics of the measurements on the
respective microparticle. Particularly note that the value of $\Delta _\Psi
A $ can be modified only by changing the microparticle state (i.e. its wave
function $\Psi ).$ Comparatively the measuring uncertainties can be modified
by improving (or worsening) the performances of the experimental techniques,
even if the state of the measured microparticle remains unchanged.

In connection with the term ''uncertainty'' it is here the place to point
out also the following remarks. In quantum mechanics a variable $A$ is
described by an adequate operator $\widehat{A}$ which (Gudder, 1979) is a
generalized stochastic (or random) quantity. The probabilistic (stochastic)
characteristics of the considered microparticle are incorporated in its wave
function $\Psi .$ Then the expected value $\left\langle A\right\rangle _\Psi 
$ and the standard deviation $\Delta _\Psi A$ appears as quantities which
are exclusively of intrinsic nature for the respective microparticle. In
such a situation a measurements must consist of a {\it statistical} {\it %
sampling} but not of a {\it solitary detection} (determination). The
respective sampling give an output-set of data on the recorder of the
measuring instrument. From the mentioned data one obtains the out-mean $%
\left\langle A\right\rangle _{OUT}$ and out-deviation $\Delta _{OUT}A=\sqrt{%
\left\langle \left( A-\left\langle A\right\rangle _{OUT}\right)
^2\right\rangle _{OUT}}$. Then it results that in fact the measuring
uncertainties must be described by means of the differences $\left\langle
A\right\rangle _{OUT}-\left\langle A\right\rangle _\Psi $ and $\Delta
_{OUT}A-\Delta _\Psi A$ but not through the quantity $\Delta _\Psi A$. (For
other comments about the measurements regarded as here see below the Sec.
IX).

The above mentioned facts prove the groundlessness of the term
''uncertainty'' in connection with the quantities $\Delta _{TE}A$ and $%
\Delta _\Psi A.$ But such a proof must be reported as a defect of TIHR.

{\it Observation}: Sometimes, particularly in old texts, the quantities $%
\Delta _{TE}A$ and $\Delta _\Psi A$ are termed ''indeterminacy'' of the
quantum variable $A.$ If such a term is viewed to denote the
''\'{n}on-deterministic'' or ''random'' character of $A$ it can be accepted
for a natural interpretation of $\Delta _\Psi A.$ So the HR given by Eqs.
(2.3)/(4.8)/(5.13) and their generalizations from Eq. (5.6) appear to be
proper for a denomination of ''indeterminacy relations''. But then it seems
strange for the respective relations to be considered as ''crucial and
fundamental formulae'' (as in TIHR conception). This because in non-quantum
branches of probabilistic sciences an entirely similar ''indeterminacy
relation'' is regarded only as a modest, and, by no means a fundamental
formula. The alluded non-quantum ''indeterminacy relation'' expresses simply
the fact that the correlation coefficient $\left( \gamma _{AB}=\left\langle
\delta A\cdot \delta B\right\rangle /\Delta A\cdot \Delta B\right) $ takes
values within the range $\left[ -1,1\right] .$ (Schilling, 1972; Gellert 
{\it et.al., }1975{\it ). }One can see that such a non-quantum relation is
also Eq. (5.23).

\subsection{The nature and the performances of the referred experiments}

As it was shown in Sec. II one of the major supports of TIHR is the
reference to experiments of ''thought'' nature. In such a reference, by
means of ''thought motivations'', the results (ideas) from known real
experiments were transplanted in a new context. But such a transplantation
seems to be inadequate for a true scientific acceptance, mainly if the new
context is practically and conceptually different from the old one. Also it
is known that the alluded acceptance must be founded only on two pieces of
resistance: (a) concrete data from the real and specially designed
experiments, and (b) correct rigorous mathematical (logical) reasonings.

One must add another less known observation about the nature of the
experiments referred by TIHR. Practically all the respective experiments are
of ''thought''-type and (Jammer, 1974, p.81) there are not known any real
experiments capable of attesting (verifying) the TIHR with a convincing
accuracy.

The above presented facts reveal the uselessness respectively the
incorrectness of the main elements ${\bf P-2.6}$ and ${\bf P-2.8}$ of TIHR.
This means that by their existence the mentioned facts evidentiate a defect
of TIHR.

The thought experiments referred by TIHR operate with the limitative
criteria included in ${\bf P-4.1}$ and consequently with Eq. (2.1). But, as
it was mentioned in Sec. V, today there are known real experiments with
super- resolutions which overstep the respective criteria. Then it is easy
to imagine some Super-Resolution-Thought-Experiments (SRTE) for which
instead of Eq. (2.1) is satisfied the SRTE relation given by Eq. (5.1). One
observes now directly that the existence of the respective SRTE relation
incriminates the assertion ${\bf P-2.12/A}$ of TIHR. Such an incrimination
must be reported also as a defect of TIHR.

\subsection{Inaccuracy of the referred theoretical formulae}

Among the main supports of TIHR we find the theoretical Eqs. (2.3)/(4.8).
But, mathematically, the respective formula is incompletely accurate because
it fails if the condition expressed by Eq. (5.8) is not satisfied. The
complete accuracy is given by the more general Eqs. (5.7) and (5.14).

The mentioned incompleteness seems to be regarded as entirely unnatural by
the TIHR partisans - e.g. when they consider the case of the variables
angular momentum $\widehat{L}_z$ and azimuthal angle $\varphi $ (see also
below Sec. VI.F.). So, instead of the failing Eqs. (2.1)/(4.8), in order to
preserve at any price the elements like ${\bf P-2.12}$ and ${\bf P-2.13}$,
the respective partisans use a strange lot of unnaturally ''adjusted''
relations. But one can easily see that the natural attitude in the alluded
case is to refer to the Eqs. (5.7)/(5.14) or, equivalently, to Eq. (5.41).
Of course that in the discussed case the respective equations degenerate
into trivial equality 0=0 which is evidently in contradiction with the
elements ${\bf P-2.12}$ and ${\bf P-2.13}$ of TIHR.

Then one finds that one way or another TIHR is incompatible with the
absolute accuracy of the referred theoretical equations: This fact
constitutes a notable defect of TIHR

\subsection{Solitary variables}

By ${\bf P-2.10/A;M}$ TIHR states that in a measurement for a solitary
variable the quantity $\Delta A$ can be taken as unlimitedly small. But if $%
\Delta A$ is identified with $\Delta _\Psi A$ the respective quantity has a
precisely defined value (dependent on the wave function $\Psi $ describing
the state of the considered microparticle) which can not be diminished in a
measurement. Then it results that ${\bf P-2.10/A;M}$ are incorrect. The
respective results points another defect of TIHR.

\subsection{Commutable variables{\bf \ }}

The main idea of TIHR about the commutable variables is asserted in ${\bf %
P-2.11/A;M.}$ It is based on the fact that in Eq. (2.3) the product of the
corresponding quantities $\Delta _\Psi A$ and $\Delta _\Psi B$ has not a
non-null lower bound. But besides the Eq. (2.3), the respective product
satisfies also the Eq. (5.12) where the term $\left| \left\langle \left[
\delta _\Psi \widehat{A},\delta _\Psi \widehat{B}\right] _{+}\right\rangle
_\Psi \right| $ can be a non-null quantity. In this respect we quote the
following example (Dumitru, 1988):

Let be a quantum microparticle moving in a two-dimensional potential well,
characterized by the potential energy $V$=0 for $0<x<a,$ $0<y<b$ and $V$=$%
\infty $ otherwise. The corresponding wave functions are

\begin{equation}
\Psi _{n_1n_2}=\frac 2{\sqrt{ab}}\sin \left( \frac{n_1\pi }ax\right) \sin
\left( \frac{n_2\pi }by\right) \left( n_1,n_2=1,2,3.....\right)  \eqnum{6.1}
\end{equation}
As two commutable variables $A$ and $B$ with $\left| \left\langle \left[
\delta _\Psi \widehat{A},\delta _\Psi \widehat{B}\right] _{+}\right\rangle
_\Psi \right| $ $\neq 0$ we consider the Cartesian coordinates $x\acute{}$
and $y\acute{}$ given by

\[
\widehat{A}=x\acute{}=x\cos \varphi +y\sin \varphi 
\]
\begin{equation}
\widehat{B}=y\acute{}=x\sin \varphi -y\cos \varphi  \eqnum{6.2}
\end{equation}
with $0<\varphi <\frac \pi 2.$ For the case pointed by Eqs. (6.1) and (6.2)
one obtains:

\[
\Delta _\Psi A=\left[ \frac{a^2}{12}\left( 1-\frac 6{\pi ^2n_1^2}\right)
\cos ^2\varphi +\frac{b^2}{12}\left( 1-\frac 6{\pi ^2n_2^2}\right) \sin
^2\varphi \right] ^{1/2} 
\]
\begin{equation}
\Delta _\Psi B=\left[ \frac{a^2}{12}\left( 1-\frac 6{\pi ^2n_1^2}\right)
\sin ^2\varphi +\frac{b^2}{12}\left( 1-\frac 6{\pi ^2n_2^2}\right) ^2\cos
^2\varphi \right] ^{1/2}  \eqnum{6.3}
\end{equation}
\[
\left| \left\langle \delta _\Psi A,\delta _\Psi B\right\rangle _{+}\right|
_\Psi =\left[ 2\sin 2\varphi \left| \frac{a^2-b^2}{12}-\frac 6{\pi ^2}\left( 
\frac 1{n_1^2}-\frac 1{n_2^2}\right) \right| \right] ^{1/2} 
\]
Then it results that in the mentioned case the Eq. (5.12) is satisfied with
non-null quantity in the right-hand term. But such a result shows that the
main idea of TIHR about the commutable variables is doubtful. So we find
another defect of TIHR.

\subsection{The ${\bf L}_z{\bf -\varphi }$ case}

The situation of the variables $L_z$ and $\varphi $ (z-component of angular
momentum and azimuthal angle) represents one of the most controverted case
in connection with TIHR.

Firstly, it was noted that the respective variables are canonically
conjugated their quantum operators being $\widehat{L}_z=-i\hbar \frac %
\partial {\partial \varphi }$ and $\widehat{\varphi }=\varphi \cdot .$ Then
it was supposed that for $L_z{\bf -\varphi }$ case the TIHR must be
applicable similarly as for other pairs of conjugated variables-e.g. for $%
p_x $ and $x$ (Cartesian momentum and coordinate). Consequently it was
expected to have the following ordinary $L_z{\bf -\varphi }$ theoretical HR

\begin{equation}
\Delta _\Psi L_z\cdot \Delta _\Psi \varphi \geq \frac \hbar 2  \eqnum{6.4}
\end{equation}
Also one knows attempts (Kompaneyets, 1966) for introducing a
thought-experimental $L_z{\bf -\varphi }$ relation of the form

\begin{equation}
\Delta _{TE}L_z\cdot \Delta _{TE}\varphi \thickapprox \hbar  \eqnum{6.5}
\end{equation}
which is similar with the $p_x-x$ version of Eq. (2.1). But note that in
fact Eq. (6.5) is only an unmasked conversion of Eq.. (2.1).

Secondly, for the $L_z{\bf -\varphi }$ case the inapplicability of TIHR in
its usual form (presented in Sec. II) was remarked. Such remarks were
signaled in a lot of debating works (Judge, 1963; Judge and Lewis, 1963;
Judge, 1964; Bouten {\it et al}, 1965; Evett and Mahmoud, 1965; Krauss,
1965, 1968; Carruthers and Nietto, 1968; Levy-Leblond, 1976; Harris and
Strauss, 1978; Hasse, 1980; Holevo, 1981; Yamada, 1982; Galitski {\it et al.,%
} 1985). The mentioned inapplicability regards mainly from the HR Eq. (6.4)
referred to some known systems such as: an atomic electron, a rotator or a
bead shifting on a ring. Note that by their properties all the respective
systems are $\varphi $-circular (or azimuthally finite). This means that for
them: (i) $\varphi \in \left[ 0,2\pi \right] ,$ (ii) the states with $%
\varphi =0$ and $\varphi =2\pi $ coincide and (iii) the states with $\varphi
\gtrsim 0$ and $\varphi \lesssim 2\pi $ are closely adjacent. Moreover for
the mentioned systems one considers only the states which are nondegenerate
in respect with $L_z$. This means that each of such state correspond to a
distinct (eigen-) value of $L_z.$ The alluded states are described by the
wave functions: 
\begin{equation}
\Psi _m\left( \varphi \right) =\frac 1{\sqrt{2\pi }}e^{im\varphi
}\,\,\,\,\,\,\,\,\,\,\,\,\,\,\,\,\,\,\,\,(m=0,\pm 1,\pm 2,......) 
\eqnum{6.6}
\end{equation}
Then, for the demarcated states, one finds $\Delta _\Psi L_z=0$, $\Delta
_\Psi \varphi =\pi /\sqrt{3}$ and (6.4) gives the absurd results $0\geq
\hbar /2.$ Such a result drives TIHR\ in a evident deadlock.

For avoiding the alluded deadlock the THR partisans advocated the following
idea: In the $L_z{\bf -\varphi }$ case the theoretical HR must not have the
ordinary form of Eq. (6.4) but an ''adjusted version'', concordant with TIHR
vision. Thus the following lot of ''adjusted $L_z{\bf -\varphi }$ HR'' was
invented:

\begin{equation}
\frac{\Delta _\Psi L_z\cdot \Delta _\Psi \varphi }{1-3\left( \Delta _\Psi
\varphi /\pi \right) ^2}\geq 0.16\hbar  \eqnum{6.7}
\end{equation}
\begin{equation}
\frac{\left( \Delta _\Psi L_z\right) ^2\cdot \left( \Delta _\Psi \varphi
\right) ^2}{1-\left( \Delta _\Psi \varphi \right) ^2}\geq \frac{\hbar ^2}4 
\eqnum{6.8}
\end{equation}
\begin{equation}
\left( \Delta _\Psi L_z\right) ^2+\left( \frac{\hbar \omega }2\right)
^2\left( \Delta _\Psi \varphi \right) ^2\geq \frac{\hbar ^2}2\left[ \left( 
\frac 9{\pi ^2}+\omega ^2\right) ^{1/2}-\frac 3{\pi ^2}\right]  \eqnum{6.9}
\end{equation}
\begin{equation}
\frac{\Delta _\Psi L_z\cdot \Delta _\Psi \varphi }{1-3\left( \Delta _\Psi
\varphi /\pi \right) ^2}\geq \hbar \frac 23\left( \frac{V_{\min }}{V_{\max }}%
\right)  \eqnum{6.10}
\end{equation}
\begin{equation}
\left( \Delta _\Psi L_z\right) ^2\cdot \left\langle \left( \delta _\Psi
f\left( \varphi \right) \right) ^2\right\rangle _\Psi \geq \frac{\hbar ^2}4%
\left\langle \frac{df}{d\varphi }\right\rangle _\Psi ^2  \eqnum{6.11}
\end{equation}
\begin{equation}
\Delta _\Psi L_z\cdot \Delta _\Psi \varphi _1\geq \frac \hbar 2  \eqnum{6.12}
\end{equation}
\begin{equation}
\Delta _\Psi L_z\cdot \Delta _\Psi \varphi \geq \frac \hbar 2\left|
\left\langle E\left( \varphi \right) \right\rangle _\Psi \right| 
\eqnum{6.13}
\end{equation}
\begin{equation}
\Delta _\Psi L_z\cdot \Delta _\Psi \varphi \geq \frac \hbar 2\left| 1-2\pi
\left| \Psi \left( 2\pi \right) \right| ^2\right|  \eqnum{6.14}
\end{equation}
In Eq. (6.9) $\omega $ is a real nonnegative parameter. In Eq. (6.10) $%
V_{\min }$ and $V_{\max }$ represents the minimum respectively the maximum
values of $V\left( \theta \right) =\int_{-\pi }^\pi \theta \left| \Psi
\left( \varphi +\theta \right) \right| ^2d\varphi ,$ where $\theta \in
\left[ -\pi ,\pi \right] .$ In Eq. (6.11) $f\left( \varphi \right) $is a
real periodic function of $\varphi $ e.g. $f\left( \varphi \right) =\sin
\varphi $ or $f\left( \varphi \right) =\cos \varphi $ and $\delta _\Psi
f=f-\left\langle f\right\rangle _\Psi .$ In Eq. (6.12) $\varphi _1=\varphi
+2\pi N,\,\Delta _\Psi \varphi _1=\left[ 2\pi ^2\left( \frac 1{12}%
+N^2-N_1^2+N-N_1\right) \right] ^{1/2}$ and $N,N_1$=two arbitrary integer
numbers with $N\neq N_1$. In Eq. (6.13) E$\left( \varphi \right) $is a
complicated expression of $\varphi .$

Connected with the Eqs. (6.7)-(6.14) and the afferent TIHR debates the
following facts are easily observed. From a subjective view, in TIHR
literature, none of the Eqs. (6.7)-(6.14) is unanimously accepted as the
true version for theoretical $L_z{\bf -\varphi }$ HR. In a objective view
the Eqs. (6.7)-(6.14) appear as a set of completely dissimilar formulae.
This because they are not mutually equivalent and each of them is applicable
only in particular and different ''circumstances''. Moreover it is doubtful
that in the cases of Eqs. (6.7)-(6.13) the respective ''circumstances''
should have in fact real physical significances. Another aspect from an
objective view is the fact that the Eqs. (6.7)-(6.13) have no correct
support in the natural (non-adjusted) mathematical formalism of quantum
mechanics. Only Eq. (6.14) has such a support through the Eq. (5.41). The
alluded observations evince clearly the persistence of TIHR deadlock as
regards the $L_z{\bf -\varphi }$ case. But in spite of the mentioned
evidence, in our days almost all of the publications seem to cultivate the
belief that the problems of $L_z{\bf -\varphi }$ case are solved by the
adjusted Eqs. (6.7)-(6.14). In the TIHR literature the mentioned belief is
often associated with a more inciting opinion. According to the respective
opinion the ordinary theoretical HR expressed by Eq. (6.4) is incorrect for
any physical system and, consequently, the respective relation must be
prohibited. Curiously, through the alluded association, TIHR partisans seem
to ignore the thought experimental Eq. (6.5). But note that the simple
removal of the respective ignorance can not solve the TIHR deadlock
regarding the $L_z{\bf -\varphi }$ case. Moreover, such a removal is
detrimental for TIHR because the Eq. (6.5) is only a conversion of Eq. (2.1)
which is an unjustified relation (see Sec. VI.B)

As regards the TIHR attitude towards the Eq. (6.4) there is another curious
ignorance/omission. In the afferent literature it is omitted any discussion
on the $L_z$ - degenerate states of the circular systems. Such a state is
associated with a set of eigenvalue of $\widehat{L}_Z$ and it is described
by a linear superpositions of eigenfunctions of $\widehat{L}_Z$ . As example
can be taken the state of a free rigid rotator with a given energy $%
E_l=\hbar ^2l(l+1)/2J$ ($l$ = orbital quantum number, $J$ = moment of
inertia). The respective state is described by the wave function 
\begin{equation}
\Psi _l\left( \theta ,\varphi \right) =\sum_{m=-l}^lC_mY_{lm}\left( \theta
,\varphi \right)  \eqnum{6.15}
\end{equation}
where $Y_{lm}\left( \theta \right) $ are the spherical functions, $l$ and $m$
denote the orbital respectively magnetic quantum numbers while $C_m$ are
arbitrary complex constants which satisfy the condition $\sum_m\left|
C_m\right| ^2=1.$ In respect with the wavefunction given in Eq. (6.15) for
the operators $\widehat{L}_Z=-i\hbar \frac \partial {\partial \varphi }$ and 
$\widehat{\varphi }=\varphi $ one obtains 
\begin{equation}
\left( \Delta _\varphi L_Z\right) ^2=\sum_m\left| C_m\right| ^2\hbar
^2m^2-\left( \sum_m\left| C_m\right| ^2\hbar m\right) ^2  \eqnum{6.16}
\end{equation}
\[
\left( \Delta _\Psi \varphi \right) ^2=\sum_m\sum_{m\acute{}}C_m^{*}C_{m%
\acute{}}\left( Y_{lm},\varphi ^2Y_{lm\acute{}}\right) 
\]
\begin{equation}
-\left[ \sum_m\sum_{m\acute{}}C_m^{*}C_{m\acute{}}\left( Y_{lm},\varphi Y_{lm%
\acute{}}\right) \right] ^2  \eqnum{6.17}
\end{equation}
With the expressions $\Delta _\Psi L_Z$ and $\Delta _\Psi \varphi $ given by
Eqs. (6.16) and (6.17) it is possible that the HR from Eq. (6.4) to be
satisfied (For more details see below the discussions about the Eq. (8.3) in
Sec. VIII). But surprisingly such a possibility was not examined by TIHR
partisans which persevere to opine that Eq. (6.4) must be prohibited as
incorrect in respect with any physical situation. We think that such an
attitude has to be considered as a defect of TIHR\ doctrine.

Contrary to the TIHR partisans opinion about the HR given by Eq. (6.4) it is
easy to see (Dumitru, 1988, 1991), that the respective relation remains
rigorously valid at least in the case of one quantum system. The respective
system is a Quantum Torsion Pendulum (QTP) oscillating around the z-axis.
Such a QTP is completely analogous with the well-known (recti)linear
oscillator.

The states of the QTP are described by the wave functions: 
\begin{equation}
\Psi _n\left( \varphi \right) =\Psi _n\left( \xi \right) =\left( \frac{%
J\omega }{\pi \hbar }\right) ^{1/4}\frac 1{\sqrt{2^nn!\sqrt{\pi }}}e^{-\frac{%
\xi ^2}2}H_n\left( \xi \right)  \eqnum{6.18}
\end{equation}
with $\xi =\varphi \cdot \sqrt{J\omega /\hbar }$, $\varphi $=azimuthal
angle, $J$=moment of inertia, $\omega $=angular frequency, $n$=0,1,2,...=the
oscillation quantum number and $H_n\left( \xi \right) =\left( -1\right)
^n\left( e^{\xi ^2}\right) \cdot \left( d^ne^{-\xi ^2}/d\xi ^n\right) $are
the Hermite polynomials.. By its properties, QTP is $\varphi $-torsional ($%
\varphi -$non-circular or azimuthally infinite). This means that for it: (i) 
$\varphi \in \left( -\infty ,\infty \right) $ , (ii) the states with $%
\varphi =0$ and $\varphi =2\pi $ do not coincide and (iii) the states with $%
\varphi \gtrsim 0$ and $\varphi \lesssim 2\pi $ are not closely adjacent.

In the case of QTP for the variables $L_z$ and $\varphi $, described by the
operators $\widehat{L}_z=-i\hbar \frac \partial {\partial \varphi }$ and $%
\widehat{\varphi }=\varphi \cdot $, one obtains the expression:

\begin{equation}
\Delta _\Psi L_z\sqrt{\hbar J\omega \left( n+\frac 12\right) }%
\,\,\,\,,\,\,\,\,\,\,\Delta _\Psi \varphi =\sqrt{\frac \hbar {J\omega }%
\left( n+\frac 12\right) }  \eqnum{6.19}
\end{equation}
With these expressions one finds that for QTP the $L_z{\bf -\varphi }$
theoretical HR is satisfied in the ordinary/common form given by Eq. (6.4).
So the existence of QTP example invalidate the above mentioned opinion of
TIHR partisans about the HR from Eq. (6.4).The alluded invalidation makes
even deeper the deadlock of TIHR with respect to the $L_z{\bf -\varphi }$
case.

All the above mentioned deadlooks of TIHR-doctrine in connection with the
pair $L_z{\bf -\varphi }$ must be reported as indubitable defects of the
respective doctrine.

\subsection{The N-${\bf \Phi }$ case}

Another case which drived TIHR in deadlock is that of pair N-${\bf \Phi }$
(number-phase) connected with the study of quantum oscillator. N{\bf \ }%
represents the quantum oscillation number, described by the operator $%
\widehat{N}=\widehat{a}^{+}\widehat{a}$ (where $\widehat{a}^{+}$ and $%
\widehat{a}$ are the known ladder operators) while $\Phi $ is taken as
variable conjugated with $N.$ Often, if the oscillator is considered as
radiative, $N$ and $\Phi $ are regarded as number respectively phase of the
radiated particles (photons or phonons). In the $\Phi $ - representation we
have

\begin{equation}
\widehat{N}=i\frac \partial {\partial \Phi }\,\,\,\,,\,\,\,\widehat{\Phi }%
=\Phi \cdot \,,\,\,\,\,\,\,\,\,\,\,\left[ \widehat{N},\widehat{\Phi }\right]
_{-}=i  \eqnum{6.20}
\end{equation}
Note that in the mentioned representation the states under the
considerations are $\Phi -$circular (in a similar way with the $\varphi $%
-circular states discussed above in connection with the $L_z{\bf -\varphi }$
case). The respective states are described by the wave functions

\begin{equation}
\Psi _N\left( \Phi \right) =\frac 1{\sqrt{2\pi }}e^{iN\Phi
}\,\,\,\,\,\,\,\,\,\,\,\,\,\,(N=0,1,2,........)  \eqnum{6.21}
\end{equation}
For the $N-\Phi $ case the TIHR doctrine requires that through the Eq. (2.3)
one should have the ordinary relation:

\begin{equation}
\Delta _\Psi N\cdot \Delta _\Psi \Phi \geq \frac 12  \eqnum{6.22}
\end{equation}
But it is easy to see that this relation is incorrect because with (6.20)
and (6.21) one obtains $\Delta _\Psi N=0$ and $\Delta _\Psi \Phi =\pi /\sqrt{%
3}.$

The incorrectness of the Eq. (6.22) derives TIHR in another deadlock. With
the aim of avoiding the respective deadlock TIHR partisans promoted the idea
to replace the Eq. (6.22) with some adjusted relations, concordant with TIHR
doctrine. So in literature (Fain and Khanin, 1965; Carruthers and Nieto,
1968; Davydov, 1973; Opatrny, 1995; Lindner et al., 1996) were promoted a
few adjusted relations such as:

\begin{equation}
\Delta _\Psi N\cdot \Delta _\Psi C\geq \frac 12\left\langle S\right\rangle 
\eqnum{6.23}
\end{equation}
\begin{equation}
\Delta _\Psi N\cdot \Delta _\Psi S\geq \frac 12\left\langle C\right\rangle 
\eqnum{6.24}
\end{equation}
\begin{equation}
\left( \Delta _\Psi \Phi \right) ^2\cdot \left[ \left( \Delta _\Psi N\acute{}%
\right) ^2+\left( \left\langle \widehat{N}\acute{}\right\rangle _\Psi +\frac %
12\right) -\left\langle \widehat{L}\acute{}_z\right\rangle \right] \geq 
\frac{\hbar ^2}4\left[ 1-\frac 3{\pi ^2}\left( \Delta _\Psi \Phi \right)
^2\right]  \eqnum{6.25}
\end{equation}
The new quantities appearing in Eqs.(6.23)-(6.25) are defined through the
relations:

\begin{equation}
\widehat{C}=\frac 12\left( \widehat{E}_{-}+\widehat{E}_{+}\right)
\,\,\,\,,\,\,\,\,\,\,\,\,\,\,\,\,\widehat{S}=\frac 1{2i}\left( \widehat{E}%
_{-}-\widehat{E}_{+}\right)  \eqnum{6.26.a}
\end{equation}
\begin{equation}
\widehat{E}_{-}\left( N+1\right) ^{-1/2}\widehat{a}\,\,\,\,\,\,\,\,\,\,,\,\,%
\,\,\,\,\,\widehat{E}_{+}=\widehat{a}^{+}\left( N+1\right) ^{-1/2} 
\eqnum{6.26.b}
\end{equation}
\begin{equation}
\widehat{N}\acute{}=\widehat{I}\widehat{L}\acute{}_z-\frac 12%
\,\,\,\,\,\,\,\,\,\,,\,\,\,\,\,\,\,\,\widehat{L}\acute{}_z=-i\hbar \frac %
\partial {\partial \Phi }  \eqnum{6.27}
\end{equation}
It is the place here to note the following observations. The replacement of
the ordinary Eq. (6.22) with the adjusted Eqs. (6.23)-(6.25) is only a
redundant mathematical operation without any true utility for physics. This
happens because for the interests of physics those of real utility are the
observables $N$ and $\Phi $ but not the above mentioned adjusted quantities $%
N\acute{},C$ or $S$. So if the interest of physics are connected on the
particles (photons or phonons) radiated by quantum oscillators the real
measuring instruments are counters respectively the polarizers. Such
instruments measure directly $N$ and $\Phi $ but not $N\acute{},C$ or $S.$
So the measuring uncertainties, appealed by TIHR as corner-stone pieces,
must regard $N$ and $\Phi $ but not $N\acute{},C$ or $S$.

The above noted observations show that the relations like Eqs. (6.23)-(6.24)
(or other adjusted formulae) do not solve the nonconformity of the pair $%
N-\Phi $ with TIHR doctrine. The respective nonconformity remains an open
question which can not be solved by means of inner elements of TIHR. This
means that the $N-\Phi $ case appears as an irrefutable defect of TIHR.

\subsection{The energy - time case}

The pair energy E - time t was and still is the subject of many debates in
the literature (Aharonov and Bohm, 1961, 1964; Fock, 1962; Alcook, 1969;
Bunge, 1970; Fujivara, 1970; Surdin, 1973; Kijovski, 1974; Bauer and Mello,
1978; Voronstov, 1981; Kobe and Aquilera-Navaro, 1994). The respective
debates originate in the following facts:

On the one hand $E$ and $t$ are considered as (canonically) conjugated
variables whoose ordinary operators $\widehat{E}=i\hbar \frac \partial {%
\partial t}$ and $\widehat{t}=t\cdot $ satisfy the commutation relation $%
\left[ \widehat{E},\widehat{t}\right] =i\hbar .$ Then for these variables
the theoretical HR expressed by Eq. (2.3) should take the ordinary form 
\begin{equation}
\Delta _\Psi E\cdot \Delta _\Psi t\geq \frac \hbar 2  \eqnum{6.28}
\end{equation}

On the other hand as $t$ is not a random but a deterministic variable $%
\Delta _\Psi t\equiv 0$ for any state (wave function). Moreover the energy
is described by the Hamiltonian operator $\widehat{H}$ and then $\Delta
_\Psi E=\Delta _\Psi H.$ Or $\Delta _\Psi H$ is a null quantity (in the case
of stationary states which are pure eigenstates of $\widehat{H}$) or a
non-null but finite quantity in the case of nonstationary states or of
stationary ones which are mixtures of eigenstates of $\widehat{H}).$ Then
one finds that in fact for the pair $E-t$ the theoretical HR given by Eqs.
(2.3)/(6.28) degenerate into the absurd result $0\geq \left( \hbar /2\right)
.$ With such a result the $E-t$ case radically deviates from the TIHR
stipulations. For avoiding the respective deviation the TIHR partisans
invented a lot of adjusted $E-t$ relations destined to replace Eq. (6.28)
and to remain concordant with the alluded stipulations. More of the
mentioned relations have the following generic form: 
\begin{equation}
\Delta _vE\cdot \Delta _vt\geq \frac \hbar 2  \eqnum{6.29}
\end{equation}
where $\Delta _vE$ and $\Delta _vt$ have various significances such as: (i) $%
\Delta _1E=$ the line breadth characterizing the decay of an excited quantum
state and $\Delta _1t=$ the duration life of the respective state (ii) $%
\Delta _2E=$ $\hbar \Delta _2\omega $ = the energetic width of a wave packet
and $\Delta _2t$ = temporal width of the packet (while Eq. (6.29) is
introduced by means of Eq.(5.35) with $x=t$ and $k=\omega =E/\hbar ).$ (iii) 
$\Delta _3E=\Delta _\Psi H$ and $\Delta _3t=\Delta _\Psi A\cdot \left(
d\left\langle A\right\rangle _\Psi /dt\right) ^{-1}$ with $A$ as an
arbitrary variable.

Other substitutions of Eq. (6.28) were adjusted by means of some strange
ideas such as

(i) to take

\begin{equation}
\Delta t=\left[ \left\langle t^2\right\rangle -\left\langle t\right\rangle
^2\right] ^{1/2}  \eqnum{6.30}
\end{equation}
with

\begin{equation}
\left\langle t^n\right\rangle =\frac{\int_{-\infty }^\infty t^n\left| \Psi
\left( q,t\right) \right| ^2dt}{\int_{-\infty }^\infty \left| \Psi \left(
q,t\right) \right| ^2dt}  \eqnum{6.31}
\end{equation}

(ii) to fabricate a time (or tempus) operator $\widehat{T}$ capable of
satisfying the commutation relation:

\begin{equation}
\left[ \widehat{T},\widehat{E}\right] =i\hbar  \eqnum{6.32}
\end{equation}
In connection with the above mentioned substitutions of Eq. (6.28) the
following major shortcomings must be notified: (i) The Eqs. (6.29)-(6.32) do
not result from the standard mathematical procedures specific for the true
theoretical HR (as presented in sections III, IV, and V), (ii) The variety
of significances for $\Delta _vE$ and $\Delta _vt$ from Eq. (6.29) still
generates persistent disputes among TIHR partisans, (iii) None of the
substitution alternatives for Eq. (6.28) was ratified until now by natural
and credible arguments.

The above notifications together with the presented observations about Eq.
(6.28) clearly disclose an important and persistent defect of TIHR.

\subsection{Bitemporal relations}

The bitemporal relation given by Eq. (5.30) facilitates the detection of
another defect of TIHR. If in the respective relation one takes $\left|
t_2-t_1\right| \rightarrow \infty $ in the TIHR vision the quantities $%
\Delta _{\Psi 1}A$ and $\Delta _{\Psi 2}B$ refer to two variables $A$ and $B$
each of them considered as solitary. In such a case the TIHR assertion ${\bf %
P-2.10/A}$ claims that both $\Delta _{\Psi 1}A$ and $\Delta _{\Psi 2}B$
should be boundlessly small quantities. But from Eq. (5.30) one can see that
such a claim is refuted in the cases when $\left( \delta _{\Psi _1}\widehat{A%
}\Psi _1,\,\delta _{\Psi _2}\widehat{A}\Psi _2\right) \neq 0$. An example of
such a case can be provided as follows.

For a QTP the time-dependent wave functions describing its states are

\begin{equation}
\Psi _{nt}=\Psi _n\left( \xi ,t\right) =\exp \left( \frac{-iE_n}\hbar
t\right) \Psi _n\left( \xi \right)  \eqnum{6.33}
\end{equation}
where $\Psi _n\left( \xi \right) $are given by Eq. (6.18) and $E_n=\hbar
\omega \left( n+\frac 12\right) .$ Then with $\widehat{A}=\widehat{L}_z$ and 
$\widehat{B}=\widehat{\varphi }$ the Eq. (5.30) becomes

\begin{equation}
\Delta _{\Psi _{nt_1}}L_z\cdot \Delta _{\Psi _{nt_2}}\varphi \geq \left|
\left( \delta _{\Psi _{nt_1}}\widehat{L}_z\Psi _{nt_1},\,\delta _{\Psi
_{nt_2}}\widehat{\varphi }\Psi _{nt_2}\right) \right|  \eqnum{6.34}
\end{equation}
with

\begin{equation}
\Delta _{\Psi _{nt_1}}L_z=\sqrt{\hbar J\omega \left( n+\frac 12\right) }%
,\,\,\,\,\Delta _{\Psi _{nt_2}}\varphi =\sqrt{\frac \hbar {J\omega }\left( n+%
\frac 12\right) }\,  \eqnum{6.35}
\end{equation}
\begin{equation}
\left( \delta _{\Psi _{nt_1}}\widehat{L}_z\Psi _{nt_1},\delta _{\Psi _{nt_2}}%
\widehat{\varphi }\Psi _{nt_2}\right) =-\frac{i\hbar }2\exp \left\{ -i\omega
\left( n+\frac 12\right) \left( t_2-t_1\right) \right\}  \eqnum{6.36}
\end{equation}

The above mentioned refutation of TIHR claimed by the bitemporal relations
given by Eqs. (5.30)/(6.34) must be notified as a clear defect of TIHR.
Various attempts (Levly-Leblond 1967; Ghanapragasam and Srinivas, 1979) of
extrapolating TIHR vision onto the relations of Eq. (5.30) type seem to us
to be without any real physical foundation.

\subsection{Multivariable relations}

In Sec. $V$ we have shown that the multivariable relations from Eqs.(5.6)
belong to the same family of theoretical formulae as the primary HR given by
Eqs. (2.2)/(2.3). For an example of such a multivariable relation we can
consider the QTP described by (6.18) and the following set of three
variables: $A_1=L_z$,\thinspace $A_2=\varphi $ and $A_3=T=L_z^2/2J=$the
kinetic energy. Then from Eq. (5.6) one obtains:

\[
\left( \Delta _\Psi L_z\right) ^2\cdot \left( \Delta _\Psi \varphi \right)
^2\cdot \left( \Delta _\Psi T\right) ^2\geq \left( \Delta _\Psi L_z\right)
^2\cdot \left| \left( \delta _\Psi \widehat{\varphi }\Psi ,\delta _\Psi 
\widehat{T}\Psi \right) \right| ^2+ 
\]

\begin{equation}
+\left( \Delta _\Psi \varphi \right) ^2\cdot \left| \left( \delta _\Psi 
\widehat{T}\Psi ,\delta _\Psi \widehat{L}_z\Psi \right) \right| ^2+\left(
\Delta _\Psi T\right) ^2\cdot \left| \left( \delta _\Psi \widehat{L}_z\Psi
,\delta _\Psi \widehat{\varphi }\Psi \right) \right| ^2-  \eqnum{6.37}
\end{equation}

\[
-2%
\mathop{\rm Re}%
\left\{ \left( \delta _\Psi \widehat{L}_z\Psi ,\delta _\Psi \widehat{\varphi 
}\Psi \right) \cdot \left( \delta _\Psi \widehat{\varphi }\Psi ,\delta _\Psi 
\widehat{T}\Psi \right) \cdot \left( \delta _\Psi \widehat{T}\Psi ,\delta
_\Psi \widehat{L}_z\right) \right\} 
\]

where $%
\mathop{\rm Re}%
f$ = real part of $f,\Delta _\Psi L_z$ respectively $\Delta _\Psi \varphi $
are given by Eq.(6.19) and

\[
\Delta _\Psi T=\frac{\hbar \omega }2\sqrt{\frac{n^2+n+1}2,}%
\,\,\,\,\,\,\,\,\,\,\,\,\,\,\,\left( \delta _\Psi \widehat{L}_z\Psi ,\delta
_\Psi \widehat{\varphi }\Psi \right) =-i\frac \hbar 2 
\]
\begin{equation}
\left( \delta _\Psi \widehat{\varphi },\delta _\Psi \widehat{T}\Psi
\,\right)
=0,\,\,\,\,\,\,\,\,\,\,\,\,\,\,\,\,\,\,\,\,\,\,\,\,\,\,\,\,\,\,\,\,\,\,%
\left( \delta _\Psi \widehat{T}\Psi ,\delta _\Psi \widehat{L}_z\Psi \right)
=0  \eqnum{6.38}
\end{equation}

The alluded relationship of the multivariable Eqs. (5.6), (6.37) with the HR
from Eqs. (2.9)/(2.10) naturally requires an argued answer to the question:
In what rapports must be the (physical) interpretation of the respective
relations with TIHR? The question was approached (Synge, 1971) from a
viewpoint of extrapolating TIHR vision. In the spirit of such a view there
was promoted the idea that a 3-variable relation of Eq. (5.6) type (obtained
by Synge on a different way) must be interpreted as describing a fundamental
interconnection among the uncertainties characterizing the simultaneous
measurements for the corresponding variables. Other scientists (Dodonov and
Man'ko, 1987), who investigate from a mathematical viewpoint the
generalizations of HR given by Eq. (2.2)/(2.3), have deliberately (and
declaratively) omitted any approach of the above mentioned question.

A careful examination of the facts shows that the above mentioned
extrapolation of TIHR is unjustifiable at least because of the following
reasons: (i) {}It cannot be sustained by real (nonfictitious) arguments
regarding the true characteristics of the measurements, (ii) As it was
pointed out in Sec. VI.A the theoretical quantities implied in Eqs. (5.6)
and (6.37) (like $\Delta _\Psi A_j$ or ($\delta _\Psi \widehat{A}_j\Psi
,\delta _\Psi A_e\Psi ))$ have not significance of measuring uncertainties.
Then it results that, although mathematically the multivariable relations
are closely related with HR, they cannot be interpreted in a manner
consonant with TIHR. But such a result must be notified as another defect of
TIHR regarded as a comprehensive doctrine.

\subsection{Thermal relations}

TIHR\ was promoted in connection with the quantum HR given by Eqs.
(2.2)/(2.3). But we have shown that, mathematically, the respective HR are
completely similar with the thermal relations from Eqs. (5.21)-(5.23).
Firstly, we shall present here some concrete examples of such thermal
relations.

So from the phenomenological (quasithermodynamic) theory of fluctuations we
can quote (Dumitru, 1974 a, 1988) the following P-V (pressure-volume) formula

\begin{equation}
\Delta _wV\cdot \Delta _wP\geq \left| \left\langle \delta _wV\delta _\omega
P\right\rangle _w\right|  \eqnum{6.39}
\end{equation}
where

\[
\Delta _wV=\sqrt{-k\overline{T}\left( \frac{\partial \overline{V}}{\partial 
\overline{p}}\right) _{\overline{T}}},\,\,\,\,\,\,\,\,\,\,\,\Delta _wP=\sqrt{%
-k\overline{T}\left( \frac{\partial \overline{p}}{\partial \overline{V}}%
\right) _{\overline{S}}}\,\,\, 
\]
\begin{equation}
\left\langle \delta _wV\delta _wP\right\rangle _w=-k\overline{T} 
\eqnum{6.40}
\end{equation}
with $\overline{A}=\left\langle A\right\rangle _w$, k = Boltzmann's constant
and S = entropy.

Also we can quote a formula of Eq. (5.23) type from classical statistical
physics. For this let us consider an ideal gas situated in a cylindrical
recipient of height b on the Earth surface. The gas is supposed to be in a
thermodynamical equilibrium state described by the canonical distribution $%
w\sim \exp \left\{ -H/kT\right\} $, with $H$ = the Hamiltonian of the
molecules. If the molecules are considered as identic point particles, we
can take $H$ of the form $H=\sum_{i=1}^{3N}\left( p_i^2/2m\right)
+\sum_{i=1}^Nmgz_i$ ($N$, $p_i$ and $z_i$ represent respectively the number,
linear momenta and altitude of the molecules; $m$ denotes the mass of a
molecule and $g$ is the gravitational acceleration). As stochastic variables
for the gas regarded as a statistical system we consider $A=H$ = Hamiltonian
and $B=Z_c$= the altitude of the centre of mass. For such variables Eq.
(5.23) transcribes as

\begin{equation}
\Delta _wH\cdot \Delta _wZ_c\geq \left| \left\langle \delta _wH\delta
_wZ_c\right\rangle _w\right| \cdot  \eqnum{6.41}
\end{equation}
The terms from this relation are given by

\begin{equation}
\left( \Delta _wH\right) ^2=k\nu RT^2\left[ \frac 52-a^2f(a)\right] 
\eqnum{6.42a}
\end{equation}
\begin{equation}
\left( \Delta _wZ_c\right) =k\frac{b^2}{\nu R}\left[ \frac 1{a^2}-f(a)\right]
\eqnum{6.42b}
\end{equation}
\begin{equation}
\left\langle \delta _wH\cdot \delta Z_c\right\rangle =kTb\left[ \left(
e^a-1\right) ^{-1}-f(a)\right]  \eqnum{6.42c}
\end{equation}
where $\nu =$ number of moles, $f(a)=e^a\left( e^a-1\right) ^{-2}$ and $%
a=\left( \mu gb/RT\right) $with $\mu =$ molar mass of the gas, $R=$
universal gas constant.

Another nontrivial exemplification of Eq. (5.22) is given by the early known
F\H{u}rth\'{}s formula from the theory of Brownian motion (F\H{u}rth, 1933).
In the respective formula $A=x$ and $B=v$ i.e. the coordinate and velocity
of a Brownian particle, whereas $\left\langle \delta _wA\delta
_wB\right\rangle _w=\left\langle \delta _wx\cdot \delta _wv\right\rangle
_w=D $ diffusion coefficient of the particle. Then the F\H{u}rth\'{}s
formula is: 
\begin{equation}
\Delta _wx\cdot \Delta _wv>D  \eqnum{6.43}
\end{equation}
Now let us return to the mentioned mathematical similarity between thermal
relations given by Eqs. (5.21)-(5.23), (6.39), (6.41), (6.43) and HR from
Eqs. (2.2)/(2.3). The respective similarity suggests an investigation of the
possible connections between the interpretation of the alluded thermal
relations respectively the TIHR. Such an investigation was approached by
some traditionalist-scientists (Frank-Kamenetsky, 1940; Bohm, 1957;
Rosenfeld, 1961 and 1962; Terletsky, 1974). Mainly, they promoted the idea
that the mentioned thermal relations must be interpreted in terms of a
macroscopic complementarity. Consequently, by extrapolation of the TIHR
statements ${\bf P-2.15}$ and ${\bf P-2.16}$, in Eq. (5.23) the variables $A$
and $B$ must be regarded as complementary (i.e. mutually exclusive) while $%
\Delta _wA$ and $\Delta _wB$ have to be interpreted as uncertainties in
simultaneous measurements. The mentioned traditionalist idea was partially
reviewed from various perspectives (Shaposhnikov, 1947; Bazarov, 1979;
Uffink and van Lith, 1999). But in the alluded reviews there are not pointed
out explicitely the facts that the quantities like DwA and DwB describe the
thermal fluctuations respectively that such fluctuations are inrinsic
properties of tehe thermodynamic systems.

Our opinion is that the alluded idea of macroscopic complementarity must be
rejected due to the following reasons: (i) The quantity $\Delta _wA$ is not
a measuring uncertainty but a parameter (standard deviation) characterizing
the thermal fluctuations. (ii) The value of $\Delta _wA$ can be modified
only by changing the inner state (i.e. the distribution function {\it w}) of
the considered system but not by means of improvements of measuring
precision. (iii) Regarded as fluctuation parameter $\Delta _wA$ can be
measured without restriction of principle. So in noise spectroscopy
(Weissmann, 1981) it is possible to measure even the ''constitutive'' (i.e.
spectral) components of $\Delta _wA.$ (iv) In classical physics the
variables characterizing a macroscopic system are not mutually exclusive
(i.e. complementary), (v) The true conception about the macroscopic
measurements does not include any reference to the mutual unavoidableness of
simultan uncertainties for two (or more) variables.

The above mentioned reasons clearly guide us to the following conclusion. In
spite of the mathematical similarity between HR and the discussed thermal
relations, TIHR can not be extended (by similitude) to the interpretation of
the respective relations. But as the mentioned mathematical similarity is of
a conceptually fundamental nature the concluded inextensibility must be
notified as a true defect of TIHR.

\subsection{The so-called macroscopic operators}

Controversies about TIHR also included several discussions regarding the
macroscopic relation given by Eq. (5.29) (see Jancel, 1973 and references).
The respective discussions were generated by the following conflicting
findings: (i) On the one hand the respective relation appears within quantum
theory and, mathematically, it is completely similar with HR expressed by
Eqs. (2.2)-(2.3). Then by extrapolating TIHR the Eq. (5.29) should be
interpreted as an interconnection of the macroscopic uncertainties $\Delta
_\rho A$ and $\Delta _\rho B$ regarding the simultaneous measurements of the
variables $A$ and $B$ afferent to a macroscopic system.

(ii) On the other hand, according to ${\bf P-2.4}$, TIHR operates with the
hypothesis that a macroscopic variable can be measured without any
uncertainty (i.e. with unbounded accuracy), irrespective of the fact that it
is measured solitary or simultaneously with other variable. Then, of course,
it is useless to speak of an interconnection between the uncertainties of
two macroscopic variables-even if, in theoretical framework, they are
described by quantum statistical operators.

To elude the mentioned conflict the TIHR partisans promoted the strange idea
to abrogate the Eq. (5.29) and to replace it by an adjusted macroscopic
formula concordant with TIHR vision. With this aim, the common operators $%
\widehat{A}$ and $\widehat{B}$ from Eq. (5.29) were substituted by the
so-called ''macroscopic operators''{\it \ }$\widehat{A}_M$ and $\widehat{B}%
_M $ which (in any representation?) can be pictured as quasi-diagonal
matrices. Then one supposes that $\left[ \widehat{A}_M,\widehat{B}_M\right]
_{-}=0$ and instead of Eq. (5.29) one obtains 
\begin{equation}
\Delta _\rho A_M\cdot \Delta _\rho B_M\geq 0  \eqnum{6.44}
\end{equation}
In this relation TIHR partisans view the fact that, simultaneously, the
uncertainties $\Delta _\rho A$ and $\Delta _\rho B$ can be arbitrarily
small. Such a view is concordant with the main concept of TIHR. Today many
scientists believe that the adjusted Eq. (6.44) solves all the troubles of
TIHR generated by the Eq. (5.29).

It is easy to remark that the mentioned belief proves to be unfounded if one
takes into account the following observations:

(i) Equation (5.29) can not be abrogated unless the entire mathematical
apparatus of quantum statistical physics is abrogated too. More exactly, the
substitution of the common operators $\widehat{A}_M$ and $\widehat{B}_M$
with macroscopic operators $\widehat{A}_M$ and $\widehat{B}_M$ is a useless
invention. This because in the practical domain of quantum statistical
physics (see for example Tyablikov, 1975) the common operators but not the
macroscopic ones are used.

(ii) The above mentioned substitution of operators does not metamorphose
automatically Eq. (5.29) into Eq. (6.44). This because, if two operators are
quasi-diagonal (in the sense required by the TIHR partisans) they can be
non-commutable. As an example in this sense we refer to a macroscopic system
formed by a large number N of independent $\frac 12$-spins (Dumitru, 1988,
1989). The hinted macroscopic variables are components $M_j$ $\left(
j=x,y,z\right) $of the magnetization $\overrightarrow{M}$. The corresponding
operators are 
\begin{equation}
\widehat{A}_j=\widehat{M}_j=\frac{\gamma \hbar }2\widehat{\sigma }_j^{\left(
1\right) }\oplus \frac{\gamma \hbar }2\widehat{\sigma }_j^{\left( 2\right)
}\oplus ......\oplus \frac{\gamma \hbar }2\widehat{\sigma }_j^{\left(
N\right) }  \eqnum{6.45}
\end{equation}
where $\gamma $ = magneto-mechanical factor, $\widehat{\sigma }_j^{\left(
n\right) }$= Pauli matrices associated with the ''n-th'' spin
(microparticle). One can see that the operators defined by Eqs. (6.45) are
quasidiagonal in the sense required for ''macroscopic operators'', but they
are not commutable among them, as we have for example $\left[ \widehat{M_x},%
\widehat{M_y}\right] =i\hbar \gamma \widehat{M_z}.$ Consequently one can say
that by the mentioned substitution of operators the Eq. (5.29) is transposed
in fact not in (6.44) but into the formula 
\begin{equation}
\Delta _\rho A_M\cdot \Delta _\rho B_M\geq \left| \left\langle \left[ 
\widehat{A}_M,\widehat{B}_M\right] _{-}\right\rangle _\rho \right| 
\eqnum{6.46}
\end{equation}
But such a formula is not helpful for TIHR if $\left\langle \left[ \widehat{A%
}_M,\widehat{B}_M\right] \right\rangle _\rho \neq 0$, as in the case of
operators defined by Eqs. (6.45).

(iii) The alluded substitution of operators does not solve the troubles of
TIHR even if the macroscopic operators are commutable. This because Eq.
(5.29) is only a truncated version of the more general Eq. (5.27). Then by
the mentioned substitution, in fact, one must consider the methamorphosis of
Eq. (5.27) which gives 
\begin{equation}
\Delta _\rho A_M\cdot \Delta _\rho B_M\geq \left| \left\langle \delta _\rho 
\widehat{A}_M\cdot \delta _\rho \widehat{B}_M\right\rangle _\rho \right| 
\eqnum{6.47}
\end{equation}
In this formula, if $\left[ \widehat{A}_M,\widehat{B}_M\right] _{-}=0$, one
obtains $\left\langle \Delta _\rho \widehat{A}_M\cdot \Delta _\rho \widehat{B%
}_M\right\rangle _\rho =\frac 12\left\langle \left[ \delta _\rho \widehat{A}%
_M,\delta _\rho \widehat{B}_M\right] _{+}\right\rangle _\rho $ i.e. a
quantity which can have a non-null value. Then it results that the
macroscopic product $\delta _\rho A_M\cdot \delta _\rho B_M$ can have a
non-null lower bound. But such a result opposes to the agreements of TIHR.

So we conclude that in fact the mentioned macroscopic operators cannot solve
the TIHR deficiencies connected with the Eq. (5.29). This means that the
respective deficiencies remain unsolved and they must be reported as another
insurmountable defect of TIHR.

\section{INDUBITABLE FAILURE\ OF\ TIHR}

A mindful examination of all the details of the facts discussed in the
previous section guide us to the following remarks:

$\underline{{\bf P-7.1}}$ : Taken together, in an ensemble, the above
presented defects incriminate and invalidate each of the main elements of
TIHR.$\blacktriangle $

\underline{${\bf P-7.2}$} : The mentioned defects are insurmountable for the
TIHR doctrine, because they cannot be avoided or refuted by means of
credible arguments from the framework of the respective doctrine.$%
\blacktriangle $

The two remarks reveal directly the indubitable failure of TIHR which now
appears as an unjustified doctrine. Then the sole reasonable attitude is to
abandon TIHR and, as a first urgency, to search the genuine significance
(interpretation) of the HR. As second urgency, probably, it is necessary a
re-evaluation of those problems in which, by its implications, TIHR persists
as a source of misconceptions and confusions.

\section{THE\ GENUINE\ SIGNIFICANCE\ OF\ H R}

A veritable search regarding the genuine significance of HR must be founded
on the true meaning of the elements implied in the introduction of the
respective relations. Then we have to take into account the following
considerations.

Firstly, we opine that thought-experimental HR of the type given by Eq.
(2.1) must be omitted from discussions. This because, as it was pointed out
(see Sec. VI.B and comments about the Eq. (5.1)), such a type of relations
has a circumstantial character dependent on the performances of the supposed
measuring-experiment. Also in the respective relations the involved
variables are not regarded as stochastic quantities such are the true
quantum variables. So the equations of (2.1) - type have not a noticeable
importance for the conceptual foundation of quantum mechanics. Moreover the
usages of such relations in various pseudo-demonstration (Tarasov, 1980)
have not a real scientific value. That is why we opine that the
thought-experimental HR of the type given by Eq. (2.1) must be completely
excluded from physics.

We resume the above opinions under the following point:

$\underline{{\bf P-8.1}}$: The thought-experimental HR like Eq. (2.1) must
be disregarded being fictitious formulae without a true physical
significance.$\blacktriangle $

As regard the theoretical HR of the kind illustrated by Eqs. (2.2)/(2.3) the
situation is completely different. The respective HR are mathematically
justified for precisely defined conditions within the theoretical framework
of quantum mechanics. This means that the physical significance
(interpretation) of the theoretical HR is a question of notifiable
importance. Now note that the mentioned HR belong to the large family of
correlation relations reviewed in Sec. V. This fact suggests that the
genuine significance (interpretation) of the theoretical HR must be
completely similar with that of the mentioned correlation relations.

We opine that the alluded suggestion must be taken into account with maximum
consideration. Then, firstly, we remark that all of the mentioned
correlation relations refer to the variables with stochastic
characteristics. Such variables are specific both for quantum and
non-quantum (i.e. classical) physical systems. Secondly, let us regard the
quantities like $\Delta _wA$ or $\left\langle \delta _wA\delta
_wB\right\rangle _w$ which appear in the corresponding correlation relations
from classical statistical physics. In our days science the respective
quantities are unanimously interpreted as {\it fluctuation parameters} of
the considered variables $A$ and $B$. Also it is clearly accepted the fact
that the respective parameters describe intrinsic properties of the viewed
systems but not some characteristics (i.e. uncertainties) of the
measurements on the respective properties. In this sense in some scientific
domains, such as noise spectroscopy (Weissman, 1981), the evaluation of the
quantities like $\Delta _wA$ is regarded as a tool for investigation of the
intrinsic properties of the physical systems. In classical conception the
description of the intrinsic properties of the physical systems is supposed
to be not amalgamated with elements regarding the measuring uncertainties.
The alluded description is made, in terms of corresponding physical
variables, within the framework of known chapters of classical physics (i.e.
mechanics, electrodynamics, optics, thermodynamics and statistical physics).
For the mentioned variables, when it is the case, the description approaches
also the fluctuations characteristics. Otherwise, in classical conception
the measuring uncertainties /errors are studied within the framework of
error analysis. Note that the respective analysis is a scientific branch
which is independent and additional with respect to the mentioned chapters
of physics (Worthing and Geffner, 1955).

The above mentioned aspects about the classical quantities $\Delta _wA$ and $%
\left\langle \delta _wA\delta _wB\right\rangle _w$ must be taken into
account for the specification of the genuine significance (interpretation)
of the quantum quantities $\Delta _\Psi A$ and $\left\langle \delta _\Psi
A\delta _\Psi B\right\rangle _\Psi $, as well as of the theoretical HR. We
think that the respective specification can be structured through the
following points:

\underline{${\bf P-8.2}$} : The quantum variables (operators) must be
regarded as stochastic quantities which admit fluctuations.$\blacktriangle $

\underline{${\bf P-8.3}$} : Acording to the usual quantum mechanics, the
time is not a stochastic variable but a deterministic quantity which does
not admit fluctuations.$\blacktriangle $

\underline{${\bf P-8.4}$} : The theoretical quantities $\Delta _\Psi
A,\,\Delta _\Psi B$,\thinspace $\left( \delta _\Psi \widehat{A}\Psi ,\delta
_\Psi \widehat{B}\Psi \,\right) $or $\left\langle \left[ \widehat{A},%
\widehat{B}\right] _{-}\right\rangle _\Psi $ must be interpreted as
parameters referring to the fluctuations regarded as intrinsic properties of
quantum microparticles.$\blacktriangle $

\underline{${\bf P-8.5}$} : The theoretical HR of the kind illustrated by
Eqs (2.2)/(2.3) must be considered through their accurate and complete forms
presented in Sec.V. In such forms they have to be interpreted as fluctuation
formulae regarding intrinsic characteristics of microparticles.$%
\blacktriangle $

\underline{${\bf P-8.6}$} : The quantities mentioned in ${\bf P-8.4}$ as
well as the theoretical HR have no connection with the description of
measuring uncertainties.$\blacktriangle $

\underline{${\bf P-8.7}$} : The description of measurements, characteristics
(e.g. of measuring uncertainties) for quantum microparticles must be made in
the frame of a scientific branch which is independent and additional with
respect to the usual quantum mechanics.$\blacktriangle $

A persuasive argumentation of these points results directly form the above
presented considerations. For ${\bf P-8.4}$ and ${\bf P-8.6}$ the respective
argumentation can be improved by the following observations: (i) the values
of the quantities discussed in ${\bf P-8.4}$ depend (through the wave
function $\Psi $) only on the considered microparticle state, (ii) the
respective values are independent of the measuring uncertainties which, for
the same measured state, can be modified (by changing the accuracy of
experimental instruments).

The above presented points ${\bf P-8.1}$ --- ${\bf P-8.6}$ can be regarded
as the main elements of a genuine reinterpretation of HR. It is easy to see
that the respective reinterpretation is completely concordant with the
working procedures of usual quantum mechanics (which, of course, include the
theoretical HR as particular formulae).

The mentioned reinterpretation assures a vision where all the defects of
TIHR presented in Sec. VI. A ---VI. L are eliminated as inadequate and
unfounded statements. So the defect discussed in Sec. VI. A. loses any sense
if one takes into account ${\bf P-8.1,\,P-8.4}$ and ${\bf P-8.6}$. Also the
defect revealed in Sec. VI. B has no value if we accept ${\bf P-8.1}$. The
facts presented in Sec. VI. C can be elucidated by considering ${\bf P-8.6.}$
For the case of solitary variables approached in Sec. VI. D a pertinent
answer is given by ${\bf P-8.4}$ and ${\bf P-8.6}$. The same points ${\bf %
P-8.4}$ and ${\bf P-8.6}$ offer a natural vision on the case of commutable
variables approached in Sec.VI. E.

The cases of the pairs $L_Z-\varphi $ and $N-\Phi $ discussed in Secs. VI. F
and VI. G can be brought in concordance with the proposed reinterpretation
of HR as follows.

In the situations described by the wave functions given by Eqs.(6.6) and
(6.21) the conditions expressed by Eqs. (5.8) are not satisfied. Then,
according to ${\bf P-8.5}$, in such situations the discussions must refer to
the complete and general correlation relations given by Eqs. (5.7) or (5.14)
but not to the restrictive formulae like Eqs. (2.3)/(5.13)/(6.4).

Note that, from a general perspective, the situations$_{\text{ }}$described
by the wave functions $\Psi $ noted in Eqs. (6.6) or (6.21) refer to the
cases which regard the uni-variable eigenstates. In such a state for two
variables A and B we have $\widehat{A}\psi =a\psi $ and $\widehat{B}\psi
\neq b\psi ${\bf \ }respectively $\Delta _\psi {}A=0$ and $0<\Delta _\psi
B\neq \infty $. But if $\left\langle \left[ \widehat{A},\widehat{B}\right]
_{-}\right\rangle ${}$_\Psi \neq 0$ for the pair $A-B$ the HR from Eq. (2.3)
and the related TIHR assertions are not applicable. It must be reminded that
an early modest notification (Davidson, 1965) of the TIHR shortcomings in
respect with the uni-variable eigenstates seem to be ignored by the TIHR
partisans.

Now we can see that the alluded inapplicability of Eq. (2.3) is generated by
the fact that for the mentioned uni-variable eigenstates the Eqs. (5.8) are
not satisfied. This because if for such states Eqs. (5.8) are satisfied {}we
must admit the following row of relations.

\[
a<B>_\Psi =a\left( \Psi ,\widehat{B}\Psi \right) =\left( \widehat{A}\Psi ,%
\widehat{B}\Psi \right) =\left( \Psi ,\widehat{A}\widehat{B}\Psi \right) = 
\]

\begin{equation}
=\left( \Psi ,\left[ \widehat{A}\widehat{B}\right] _{-}\Psi \right) +\left(
\Psi ,\widehat{B}\widehat{A}\Psi \right) =\left\langle \left[ \widehat{A}%
\widehat{B}\right] _{-}\right\rangle _\Psi +<B>_\Psi \cdot a  \eqnum{8.1}
\end{equation}
i.e. the absurd result $a<B>_\Psi =\left\langle \left[ \widehat{A},\widehat{B%
}\right] _{\_}\right\rangle _\Psi +a\left\langle B\right\rangle _\Psi $ with 
$\left\langle \left[ A,B\right] _{-}\right\rangle _\Psi \neq 0$.

Add here the fact that for the discussed states instead of Eq. (2.3) the
more general Eq. (5.7) remain valid (in the trivial form 0 = 0).

It is quite evidently that the situations of uni-variable eigenstates come
into the normality if $\Delta _\Psi A$ and $\Delta _\Psi B$ are regarded as
parameters which describe the quantum fluctuations. Then in such situations
A and B have respectively have not fluctuations (i.e. stochastic
characteristics).

The situation described by the wave function given by Eqs. (6.15) must be
discussed separately. Firstly it necessitates a confrontation with the
conditions expressed by Eqs. (5.8). In this sense we note that for the
respective situation one obtains the relation 
\[
\left( \hat{L}_Z\Psi _l,\hat{\varphi}\Psi _l\right) -\left( \Psi _l,\widehat{%
L}_z\hat{\varphi}\Psi _l\right) = 
\]
\begin{equation}
=i\hbar \left\{ 1+2%
\mathop{\rm Im}%
\left[ \sum_m\sum_{m\acute{}}C_m^{+}C_{m\acute{}}m\left( Y_{lm},\hat{\varphi}%
Y_{lm\acute{}}\right) \right] \right\}  \eqnum{8.2}
\end{equation}
Where $%
\mathop{\rm Im}%
\alpha $ denote the imaginary part of the complex quantity $\alpha $. Then
one observes that in the cases when the right hand term in Eq. (8.2) is null
the variables $L_Z$ and $\varphi $ satisfy the Eqs.s (5.8). In such cases
the Eqs. (2.3)/(5.13)/(6.4) are applicable. In other cases (when the
mentioned term from Eq. (8.2) is non-null) the Eqs. (5.8) are infringed and
for the pair $L_Z-\varphi $ it must apply only the Eqs. (5.7) or (5.14).

Note that the $L_Z-\varphi $ case in the situation described by wave
functions from Eqs.(6.15) can be approached also by using the Fourier
analysis procedures. So for the mentioned case and situations, similarly
with Eq. (5.41), one obtains 
\begin{equation}
\Delta _\Psi L_Z\cdot \Delta _\Psi \varphi \geq \left| 
\mathop{\rm Im}%
\left\{ \sum_m\sum_{m\acute{}}C_m^{*}C_{m^{\prime }}\cdot m\left(
Y_{lm},\varphi Y_{lm^{\prime }}\right) \right\} \right|  \eqnum{8.3}
\end{equation}

As regards the case of QTP described by the wave functions given by Eqs.
(6.18) we note the following observations. In such a case the variables $L_Z$
and $\varphi $ satisfy the conditions expressed by Eqs. (5.8). Consequently
for the respective variables are applicable the Eqs. (2.3)/(5.13)(6.4). But
the mentioned equations must be considered as resulting from more general
Eqs. (5.7) or (5.14) which are referring to the quantum fluctuations.

The problems with the pair energy-time mentioned in Sec. VI. H become
senseless if it is accepted ${\bf P-8.3.}$ According to ${\bf P-8.5}$ the
relations mentioned in sections VI. I and VI. J become simple
generalizations of he theoretical HR without interpretational shortcoming.
The relations discussed in Sec. VI. K and VI. L are nothing but macroscopic
similars of the quantum theoretical HR. Also the respective relations do not
imply any interpretational shortcoming. Moreover, the so-called macroscopic
operators discussed in Sec. VI. L appear as pure inventions without any
physical utility or significance.

${\bf \blacklozenge }${\bf A reply addendum regarding the L}$_z{\bf -\varphi 
}${\bf \ pair}

Our first opinions about the $L_z-\varphi $ pair in connection with TIHR
were presented in earlier works (Dumitru 1977, 1980). Perhaps the respective
presentations were more modest and less complete - e.g. we did not use at
all the arguments resulting from the above mentioned examples of $L_z$ -
degenerate states or of QTP. Newertheless, we think that the alluded
opinions were correct in their essence. However, in a review (Schroeck Jr.,
1982) the respective opinions were judged as being erroneous. In this
addendum, by using some of the above discussed facts, we wish to reply to
the mentioned judgements.

The main error reproached by Prof. Schroeck to us is: ''most of the results
stated concerning angular momentum and angle operators (including the
supposed canonical commutation relations) are false, this being a
consequence of not using Rieman-Stieljes integration theory which is
necessitated since the angle function has a jump discontinuity''. In order
to answer to this reproach we resort to the following specifications: (i)
One can see that the respective reproach is founded, in fact, on the idea
that the variable $\varphi $ has a jump (of magnitude 2$\pi $ at $\varphi =0$
or, equivalently at $\varphi =2\pi )$ and, consequently, on the commutation
relation is $\left[ \widehat{L}_z,\widehat{\varphi }\right] _{-}=-i\hbar
+i\hbar 2\pi \delta $ (where $\delta $ = Dirac delta function at the
boundary $\varphi =0$ or $\varphi =2\pi ).$ Note that the respective idea
(confessed explicitly to us by Prof. Schroeck in two letters dated September
16, 1981 and April 2, 1982) can also be found in most TIHR publications.
(ii) The mentioned idea refers, in fact, only to the systems which are $%
\varphi $ - circular and non-degenerate in respect with $L_z$ (defined in
the sense precised above in Sec. VI. F). But, strangely, it is associated
with the supposition that the range of $\varphi $ is the whole domain $%
\left( -\infty ,\infty \right) $, but not the finite interval $\left[ 0,2\pi
\right] $. (iii) Here is the place to add also the cases presented in Sec.
VI. F of QTP and of the $L_z$ - degenerate states (the last ones for the
situations with non-null term in the right hand side of Eq. (8.2)). For the
respective cases we must consider, another commutation relation, namely $%
\left[ \widehat{L}_z,\widehat{\varphi }\right] =-i\hbar $. (iv) Then in the
spirit of the mentioned idea for the same pair of variables $L_z-\varphi $
one must tolerate two completely dissimilar commutation relations: $\left[ 
\widehat{L}_z,\widehat{\varphi }\right] =-i\hbar +i\hbar 2\pi \delta $ and $%
\left[ \widehat{L}_z,\widehat{\varphi }\right] =-i\hbar .$ But such a
toleration seems to be senseless and without any real (physical)
substantion. (v) Our opinion about the $\widehat{L}_z-\widehat{\varphi }$
pair remains, as it was announced in previous works, and argued with more
details in the present paper. It is founded on the necessity to approach in
a unique manner all the alluded cases. In essence we think that, in all the
respective cases, for the $L_z-\varphi $ pair we must have an unique
commutation relation, namely $\left[ \widehat{L}_z,\widehat{\varphi }\right]
_{-}=-i\hbar .$ The implementation of the respective relation in the
mentioned cases for obtaining theoretical formulae of HR-type must be made
by taking into account the natural (physical) range of $\varphi $ as well as
the fulfillment of the Eqs. (5.8).

The ensemble of the above noted specifications proves as unfounded the
reproaches of Professor F.E. Schroeck Jr. regarding our opinions about the $%
L_z-\varphi $ pair.$\blacklozenge $

\[
\ast ** 
\]

The facts presented in this section show that all the problems directly
connected with the interpretation of HR can be solved by the here-proposed
genuine reinterpretation of the respective relations. But, as it is known,
TIHR generated also disputes about the topics which are adjacent or
additional with respect to the alluded problems. Several such topics will be
briefly approached in next sections.

\section{A RECONSIDERATION\ REGARDING THE\ MEASUREMENTS}

As it was mentioned in Sec. II the story of HR started with the primary
questions regarding the measuring uncertainties. During the years the
respective questions and more generally the description of the measurements
generated a large number of studies (a good list of references, in this
sense, can be obtained from the works: Yanase et al., 1978; Braginsky and
Khalili 1992; Bush et. al., 1996; Hay and Peres, 1998; Sturzu, 1999 and
surely from the bibliographical publications indicated in the end of Sec.
I). It is surprising to see that many of the above alluded studies are
contaminated one way or another by ideas pertaining to the TIHR doctrine.
After the above exposed argumentation against TIHR, here we wish to present
a few elements of a somewhat new reconsideration of the problems regarding
the description of measurements (including of measuring uncertainties). We
think that, even modestly limited, such a reconsideration can be of
non-trivial interest for our days\'{}science. This because we agree with the
opinion (Primas and M\H{u}ller - Herold, 1978) that, in fact, ''there exists
not yet a fundamental theory of actual measuring instruments''.

Firstly, we note that, in our opinion, the questions ${\bf P-2.1-P-2.2}$ are
of real significance for the studies of the physical systems. This fact is
due to the essential role of measurements (i.e. of quantitatively evaluated
experiments) for the mentioned studies. Moreover, we think that, in
principle, the alluded role appears both in quantum and non-quantum physics.

Then in our announced reconsideration we must try to search for natural
answer to the questions.. ${\bf P-2.1-P-2.2}$ as well as to some other (more
or less) directly connected problems. For such a purpose we shall note the
remarks under the following points:

\underline{${\bf P-9.1}$} : As a rule, all the measurements, both from
macroscopic and microscopic physics, are confronted with measuring
uncertainties.$\blacktriangle $

\underline{${\bf P-9.2}$} : The respective uncertainties are generated by
various factors. Among such factors the most important ones seem to be the
intrinsic fluctuations within the experimental devices and the measuring
perturbations (due to the interactions of the respective devices with the
measured systems).$\blacktriangle $

\underline{${\bf P-9.3}$} : A quantitative description of the measuring
uncertainties must be made in the framework of an authentic theory of
measurements. The respective theory must be independent and additional with
respect to the traditional chapters of physics (which describe the intrinsic
properties of physical systems).$\blacktriangle $

\underline{${\bf P-9.4}$} : The measurement of a stochastic variable should
not be reduced to a sole detection. It must be regarded and managed as a 
{\it statistical sampling} (i.e. as a statistical ensemble of detections).
Therefore, for such a variable, the finding of a single value from a sole
detection does not mean the collapse of the corresponding stochastic
characteristics (described by a wave function or by a probability density).$%
\blacktriangle $

\underline{${\bf P-9.5}$} : In the spirit of the above remark the measuring
uncertainties of stochastic variables must be described in terms of
quantities connected with the afferent statistical sampling but not with
solitary detections.$\blacktriangle $

\underline{${\bf P-9.6}$} : As regards the above alluded theory of
measurements we agree with the idea (Bunge, 1977b) that it must include some
specific elements, but not only generic-universal aspects. This because
every experimental apparatus, used in real measurements, has a well-defined
level of performance and a restricted class of utilizations - i.e. it is not
a generic-universal (for all-purpose) device.$\blacktriangle $

\underline{${\bf P-9.7}$} : Togheter with the mentioned agreement we opine
that the measurements theory can include also some elements/ideas with
generic-universal characteristics. One such characteristic is connected with
the fact that, in essence, every measurement can be regarded as an
acquisition of some information about the measured system.$\blacktriangle $

In the spirit of the latter remark we think that from a generic-universal
viewpoint a measurement can be described as a process of information
transmission, from the measured system to the receiver (recorder or
observer). In such a view, the measuring apparatus can be represented as a
channel for information transmission, whereas the measuring uncertainties
can be pictured as an alteration of the processed information. Such
informational approach is applicable for measurements on both macroscopic
and microscopic (quantum) systems. Also the mentioned approach does not
contradict to the idea of specificity as regards the measurements theory.
The respective specificity is implemented in theory by the concrete
models/descriptions of the measured system (information source), of the
measuring apparatus(transmission channel), and of the recorder/observer
(information receiver).

For an illustration of the above ideas let us refer to the description of
the measurement of a single stochastic variable $x$ having a continuous
spectrum of values within the range $\left( -\infty ,\infty \right) $. The
respective variable can be of classical kind (such are the macroscopic
quantities discussed in connection with Eqs. (5.19)-(5.23)) or of a quantum
nature (e.g. a Cartesian coordinate of a microparticle).

The alluded measurement, being regarded as a statistical sampling, its usual
task is to find certain global probabilistic parameters of $x$ such as:
mean/expected value, standard deviation, or even higher order moments. But
the respective parameters are evaluated by means of the elementary
probability $dP=w\left( x\right) dx$ of finding the value of $x$ within the
infinitesimal range $\left( x,x+dx\right) .$ Here $w\left( x\right) $
denotes the corresponding probability density. Then the mentioned task can
be connected directly to $w\left( x\right) .$

Related to the measured system\'{}s own properties, $w\left( x\right) $ has
an IN (input) expression $w_{IN}\left( x\right) $.So viewed $w_{IN}\left(
x\right) $ is assimilable with: (a) a usual distribution from classical
statistical physics - in the case of a macroscopic system, respectively (b)
the quantity $\left| \Psi \left( x\right) \right| ^2;$($\Psi \left( x\right) 
$ = the corresponding wave function) {}- in the case of a quantum
microparticle.

The fact that the measuring apparatus distorts (alters) the information
about the values of $x$ means that the respective apparatus records an OUT
(output) probability density $w_{OUT}\left( x\right) $ which generally
differs from $w_{IN}\left( x\right) $. So, with respect to the measuring
process, $w_{IN}\left( x\right) $ and $w_{OUT}\left( x\right) $ describe the
input respectively the output information. Then it results that the
measuring uncertainties (alterations of the processed information) must be
described in terms of several quantities depending on both $w_{OUT}\left(
x\right) $ and $w_{IN}\left( x\right) .$

A description of the mentioned kind can be obtained, for instance, if one
uses the following mean values: 
\begin{equation}
\left\langle f\right\rangle _A=\int_{-\infty }^\infty f\left( x\right)
\,w_A\left( x\right) dx,\,\,\,\,\left( A=IN;\,OUT\right)  \eqnum{9.1}
\end{equation}
with $f\left( x\right) $ = an arbitrary function of $x$. Then a possible
quantitative evaluation of the measuring disturbances can be made in terms
of both the following parameters:

(i) the {\it mean value uncertainty }given by 
\begin{equation}
\delta \left( \left\langle x\right\rangle \right) =\left\langle
x\right\rangle _{OUT}-\left\langle x\right\rangle _{IN}  \eqnum{9.2}
\end{equation}

(ii) the {\it standard deviation uncertainty }defined as 
\begin{equation}
\delta \left( \Delta x\right) =\Delta _{OUT}x-\Delta _{IN}x  \eqnum{9.3}
\end{equation}
where $\Delta _Ax=\left[ \left\langle x^2\right\rangle _A-\left\langle
x_A\right\rangle ^2\right] ^{1/2},\,\,\,\,\,\,(A=IN;OUT).$

The mentioned evaluation can be enriched by also using the higher order
probabilistic moments/correlations (in the sense discussed by Dumitru
(1999)).

Another evaluation of the measuring uncertainties can be made by means of
the {\it informational entropy uncertainty } 
\begin{equation}
\delta H=H_{OUT}-H_{IN}  \eqnum{9.4}
\end{equation}
Here the informational entropies $H_A\,\,(A=IN;\,OUT)\,$are defined by 
\begin{equation}
H_A=-\int_{-\infty }^\infty w_A\left( x\right) \ln \left[ 1_xw_A\left(
x\right) \right] dx  \eqnum{9.5}
\end{equation}
where $1_x$ = the unit of the physical variable $x.$

Due to the fact that, in the present considerations, a measurement is
regarded as a statistical sampling, the parameters defined by Eqs.
(9.2)-(9.4) can be called {\it statistical uncertainties.}

The uncertainty parameters introduced by Eqs. (9.2)-(9.4) can be detailed if
one takes into account more elements regarding the characteristics of the
measuring apparatus and/or of the measured system. So we can refer to an
apparatus modeled as an (information) transmission channel with stationary
and linear characteristics.. Then we can write 
\begin{equation}
w_{OUT}\left( x\right) =\int_{-\infty }^\infty G\left( x,x\acute{}\right)
\,w_{IN}\left( x\acute{}\right) dx\acute{}  \eqnum{9.6}
\end{equation}
where the kernel $G\left( x,x\acute{}\right) $ must satisfy the
normalization conditions 
\begin{equation}
\int_{-\infty }^\infty G\left( x,x\acute{}\right) dx=\int_{-\infty }^\infty
G\left( x,x\acute{}\right) dx\acute{}=1  \eqnum{9.7}
\end{equation}
The measurement is ideal or real (non-ideal) in the cases when $%
w_{OUT}\left( x\right) =w_{IN}\left( x\right) $ respectively $w_{OUT}\left(
x\right) \neq w_{IN}\left( x\right) $. In the above model such cases appear
if we take $G\left( x,x\acute{}\right) =\delta \left( x-x\acute{}\right) $
respectively $G\left( x,x\acute{}\right) \neq \delta \left( x-x\acute{}%
\right) $, with $\delta \left( x-x\acute{}\right) $ as Dirac $\delta $
function of $x-x\acute{}.$ Then by using Eqs. (9.4)-(9.6) and by taking into
account the relation $\ln y\leq y-1,\left( y>0\right) ,$ one obtains 
\[
\delta H=-\int_{-\infty }^\infty dx\int_{-\infty }^\infty dx\acute{}G\left(
x,x\acute{}\right) \,w_{IN}\left( x\acute{}\right) \ln \left[ \frac{%
w_{OUT}\left( x\right) }{w_{IN}\left( x\acute{}\right) }\right] \geq 
\]
\begin{equation}
\geq -\int_{-\infty }^\infty dx\int_{-\infty }^\infty dx\acute{}G\left( x,x%
\acute{}\right) \,w_{IN}\left( x\acute{}\right) \left[ \frac{w_{OUT}\left(
x\right) }{w_{IN}\left( x\acute{}\right) }-1\right] =0  \eqnum{9.8}
\end{equation}
So we find 
\begin{equation}
\delta H=H_{OUT}-H_{IN}\geq 0  \eqnum{9.9}
\end{equation}
This relation shows that for the processed information (measurement of $x$)
the {\it entropy H} at the recorder is greater (or at least equal) than at
the measured system. In terms of the entropy uncertainty we can speak of
ideal respectively real (non-ideal) measurement as $\delta H=0$ or $\delta
H\neq 0.$

We can detail even more the above ideas by the following example. Let us
consider $x$ as coordinate of a rectilinear quantum oscillator situated in
its lowest energy state. Such a state is described by the Gaussian wave
function 
\begin{equation}
\Psi \left( x\right) =\left( \sqrt{2\pi }\sigma _{IN}\right) ^{-1/2}\exp
\left\{ -\frac{x^2}{4\sigma _{IN}^2}\right\}  \eqnum{9.10}
\end{equation}
where $\sigma _{IN}=\hbar /2m\omega $ with $m$ and $\omega $ denoting the
mass respectively the angular frequency of the oscillator.. Then $%
w_{IN}\left( x\right) $ is given by 
\begin{equation}
w_{IN}\left( x\right) =\left| \Psi \left( x\right) \right| ^2=\left( \sqrt{%
2\pi }\sigma _{IN}\right) ^{-1}\exp \left\{ -\frac{x^2}{2\sigma _{IN}^2}%
\right\}  \eqnum{9.11}
\end{equation}
The $IN$ - values (calculated with $w_{IN}\left( x\right) $) of mean and
standard deviation of $x$ are 
\begin{equation}
\left\langle x\right\rangle _{IN}=0,\,\,\,\,\,\,\,\,\Delta _{IN}x=\sigma
_{IN}=\sqrt{\frac \hbar {2m\omega }}  \eqnum{9.12}
\end{equation}
If the errors induced by the measuring device are supposed to be small, the
kernel $G\left( x,x\acute{}\right) $ can be taken of the Gaussian form: 
\begin{equation}
G\left( x,x\acute{}\right) =\frac 1{\sigma _D\sqrt{2\pi }}\exp \left\{ -%
\frac{\left( x-\varepsilon -x\acute{}\right) }{2\sigma _D^2}\right\} 
\eqnum{9.13}
\end{equation}
Here $\varepsilon $ and $\sigma _D$ denote the {\it precision indices} of
the device. So an ideal, respectively a real measurement correspond to $%
\varepsilon \rightarrow 0$ and $\sigma _D\rightarrow 0$ (when $G\left( x,x%
\acute{}\right) \rightarrow \delta \left( x-x\acute{}\right) )$ respectively
to $\varepsilon \neq 0$ and $\sigma _D\neq 0.$ By introducing Eqs. (9.13) in
(9.6) one finds 
\begin{equation}
w_{OUT}\left( x\right) =\frac 1{\sigma _{OUT}\sqrt{2\pi }}\exp \left\{ -%
\frac{\left( x-\varepsilon \right) ^2}{2\sigma _{OUT}^2}\right\} 
\eqnum{9.14}
\end{equation}
with 
\begin{equation}
\sigma _{OUT}^2=\sigma _{IN}^2+\sigma _D^2  \eqnum{9.15}
\end{equation}
For the OUT-expressions (calculated with $w_{OUT}\left( x\right) $) of the
mean value respectively of standard deviation of $x$ one obtains 
\begin{equation}
\left\langle x\right\rangle _{OUT}=\varepsilon  \eqnum{9.16}
\end{equation}
\begin{equation}
\Delta _{OUT}x=\sigma _{OUT}=\sqrt{\left( \Delta _{IN}x\right) ^2+\sigma _D^2%
}  \eqnum{9.17}
\end{equation}
Then the uncertainties defined by Eqs. (9.2) and (9.3) become 
\begin{equation}
\delta \left( \left\langle x\right\rangle \right) =\varepsilon  \eqnum{9.18}
\end{equation}
\begin{equation}
\delta \left( \Delta x\right) =\Delta _{IN}x\left\{ \sqrt{1+\left( \frac{%
\sigma _D}{\Delta _{IN}x}\right) ^2}-1\right\}  \eqnum{9.19}
\end{equation}
In the same circumstances for the uncertainty given by Eq. (9.4) of the
informational entropy regarding $x$ one finds 
\begin{equation}
\delta H_x=\ln \left( \frac{\sigma _{OUT}}{\sigma _{IN}}\right) =\ln \sqrt{%
1+\left( \frac{\sigma _D}{\Delta _{IN}x}\right) ^2}  \eqnum{9.20}
\end{equation}
Now we can extend our discussion for the measurement of the momentum $p$ in
the case of the same quantum oscillator described by the wave function from
Eq. (9.10). As it is known the respective wave function is given in $x$
-representation. But it can be transcribed in $p$-representation in the
form: 
\begin{equation}
\Psi \left( p\right) =\left( \sqrt{2\pi }\mu _{IN}\right) ^{-1/2}\exp
\left\{ -\frac{p^2}{4\mu _{IN}}\right\}  \eqnum{9.21}
\end{equation}
with $\mu _{IN}=\left( \hbar m\omega /2\right) .$ So all the above
considerations can be transcribed from a $x$-variable version in a $p$%
-variable form.

Then the $IN$- expressions for the probability distribution respectively for
mean value and standard deviation of $p$ are 
\begin{equation}
w_{IN}\left( p\right) =\left| \Psi \left( p\right) \right| ^2=\left( \sqrt{%
2\pi }\mu _{IN}\right) ^{-1}\exp \left\{ -\frac{p^2}{2\mu _{IN}^2}\right\} 
\eqnum{9.22}
\end{equation}
\begin{equation}
\left\langle p\right\rangle _{IN}=0  \eqnum{9.23}
\end{equation}
\begin{equation}
\Delta _{IN}p=\mu _{IN}=\sqrt{\frac{\hbar m\omega }2}  \eqnum{9.24}
\end{equation}
If the $p$-measuring device is also characterized by small errors it can be
described by the kernel 
\begin{equation}
G\left( p,p^{\prime }\right) =\left( \sqrt{2\pi }\mu _D\right) ^{-1}\exp
\left\{ -\frac{\left( p-\eta -p^{\prime }\right) }{2\mu _D^2}\right\} 
\eqnum{9.25}
\end{equation}
with $\eta $ and $\mu _D$ as precision indices of the device. Similarly with
the $x$-variable case for the momentum $p$ one finds that the distribution $%
w_{OUT}\left( p\right) $characterizing the output of the measuring process
is given by 
\begin{equation}
w_{OUT}\left( p\right) =\left( \sqrt{2\pi }\cdot \mu _{OUT}\right) ^{-1}\exp
\left\{ -\frac{\left( p-\eta \right) ^2}{2\mu _{OUT}^2}\right\}  \eqnum{9.26}
\end{equation}
where 
\begin{equation}
\mu _{OUT}^2=\mu _{IN}^2+\mu _D^2  \eqnum{9.28}
\end{equation}

Then the $OUT$-expression of the mean value and standard deviation for $p$
are 
\begin{equation}
\left\langle p\right\rangle _{OUT}=\eta  \eqnum{9.29}
\end{equation}
\begin{equation}
\Delta _{OUT}p=\mu _{OUT}=\sqrt{\left( \Delta _{IN}p\right) ^2+\mu _D^2} 
\eqnum{9.30}
\end{equation}

Correspondingly, the uncertainties of the mean value and standard deviation
of $p$ are 
\begin{equation}
\delta \left( \left\langle p\right\rangle \right) =\eta  \eqnum{9.31}
\end{equation}
\begin{equation}
\delta \left( \Delta p\right) =\Delta _{IN}p\left\{ \sqrt{1+\left( \frac{\mu
_D}{\Delta _{IN}p}\right) ^2}-1\right\}  \eqnum{9.32}
\end{equation}
Also for the uncertainty regarding the informational entropy for the $p$
variable one obtains 
\begin{equation}
\delta H_p=\ln \left( \frac{\mu _{OUT}}{\mu _{IN}}\right) =\ln \sqrt{%
1+\left( \frac{\mu _D}{\Delta _{IN}p}\right) ^2}  \eqnum{9.33}
\end{equation}

Now, the above presented statistical uncertainties for $x$ and $p$ can be
regarded together for a possible comparison with some supposed ideas from
the TIHR doctrine. For such a purpose we consider the following products: 
\begin{equation}
\delta \left( \left\langle x\right\rangle \right) \cdot \delta \left(
\left\langle p\right\rangle \right) =\varepsilon \cdot \eta  \eqnum{9.34}
\end{equation}
\[
\delta \left( \Delta x\right) \cdot \delta \left( \Delta p\right) =\Delta
_{IN}x\cdot \Delta _{IN}p\cdot 
\]
\begin{equation}
x\left\{ \sqrt{1+\left( \frac{\sigma _D}{\Delta _{IN}x}\right) ^2}-1\right\}
\cdot \left\{ \sqrt{1+\left( \frac{\mu _D}{\Delta _{IN}p}\right) ^2}%
-1\right\}  \eqnum{9.35}
\end{equation}
\begin{equation}
\delta H_x\cdot \delta H_p=\ln \sqrt{1+\left( \frac{\sigma _D}{\Delta _{IN}x}%
\right) ^2}\cdot \ln \sqrt{1+\left( \frac{\mu _D}{\Delta _{IN}p}\right) ^2} 
\eqnum{9.36}
\end{equation}

As we have discussed in sections II and VI, TIHR supposes that the product
of the uncertainties for $x$ and $p$ has a non-null inferior limit. The
respective limit is expressed only in terms of the fundamental constant $%
\hbar $, and it is completely independent of certain characteristics
regarding the measuring device. Comparatively, from Eqs. (9.34)-(9.36) it
results that the products of the statistical uncertainties for $x$ and $p$
are directly dependent on the precision parameters $\varepsilon ,\eta
,\sigma _D$ and $\mu _D$ of the measuring devices. If all the respective
parameters are null (case of ideal measurements) the mentioned products are
also null. This means that the products of the mentioned statistical
uncertainties for $x$ and $p$ have not a non-null inferior limit.

Now note that, in the here-proposed model for describing the measurements,
the $x$ - and $p$ - devices are considered as completely independent. In
principle, the respective devices can be coupled in a more complex $x$ \& $p$
- instrument. A theoretical model for the description of measurements with
such an instrument can be obtained only by using a set of adequately
justified hypotheses. In such a model probably that the above presented
kernels $G\left( x,x\acute{}\right) $ and $G\left( p,p\acute{}\right) $must
be regarded as resulting from a more complex quantity dependent on both
pairs $x-x\acute{}$ and $p-p\acute{}$. But a $x-p$ couplage of the mentioned
kind still requires further investigations. Then, probably, the problem of
the product of adequate $x$-and $p$-uncertainties will be also discussed.

\section{THE\ SIMILAR\ STOCHASTIC\ SIGNIFICANCES OF\ PLANCK'S AND\
BOLTZMANN'S CONSTANTS}

In sectionsV, VI, and VII we have argued that the theoretical HR from
quantum mechanics given by Eqs. (2.2)-(2.3) have authentic nonquantum
analogs. But the mentioned HR are commonly associated with Planck\'{}s
constant $\hbar .$ Then there naturally arises the question whether $\hbar $
also has an authentic analog in nonquantum physics. Now, in this section we
shall present a lot of elements which reveal that the answer to the above
question is affirmative. The alluded analog of $\hbar $ is shown to be the
Boltzmann\'{}s constant k. The viewed analogy is given by the fact that $%
\hbar $ and k have similar roles of generic indicators of the onefold
stochasticity (randomness) for well-specified classes of physical systems
(i.e. for individual quantum microparticles and macroscopic nonquantum
systems, respectively).

A physical system is considered to have stochastic respectively
nonstochastic characteristics depending on the probabilistic nature of its
specific variables. For a system, the level (degree) of stochasticity
depends on the frame (approach) in which it is studied. So, for a
macroscopic system consisting of a large ensemble of molecules, the
stochasticity is significant in statistical physics approach but it is
completely unimportant in the frame of continuous mechanics or of
thermodynamics.. Also, in the case of a microparticle of atomic size, the
stochastic characteristics are essential from a quantum mechanics view, but
they are negligible from a classical mechanics perspective. Of course, the
level of stochasticity can be described by means of certain fluctuation
quantities such as the ones defined/implied in Eqs. (4.3), (5.4), (5.6),
(5.15), (5.20) and (5.24). But the respective quantities take particular
expressions (and values) for diverse variables or various systems.
Therefore, they cannot be considered as generic indicators of stochasticity,
i.e. as parameters indicating generically the stochasticity level for a set
or variables or for a whole class of systems. Below we shall show that the
roles of such generic indicators of stochasticity are played, in similar
ways, by the constants k and $\hbar $ for the macroscopic nonquantum systems
and individual quantum microparticles, respectively.

Firstly, let us discuss the alluded role for k with respect to the
macroscopic systems. If such a system is studied in the framework of
phenomenological (quasithermodynamic) theory of fluctuations (Munster, 1960,
1969; Dumitru, 1974a; Landau and Lifschitz, 1984), its microscopic-molecular
structure is completely neglected. Also, its specific variables are global
macroscopic quantities regarded as real stochastic variables with continuous
spectra of values. For such a system, in the mentioned approach, the
fluctuations of the variables as pressure $P$ and volume $V$ are described
by the quantities $\Delta _wV,\,\Delta _wP$ and $\left\langle \delta
_wV\delta _wP\right\rangle _w$ given in Eq. (6.40). More generally, for the
same system we can consider two arbitrary variables $A=A\left( X_j\right) $%
and $B=B\left( X_j\right) $regarded as functions of certain independent
variables $X_j\,\left( j=1,2,....,n\right) .$ Then the correlation $%
\left\langle \delta _wA\delta _wB\right\rangle _w$ describing the
fluctuations of $A$ and $B$ is given by 
\begin{equation}
\left\langle \delta _wA\delta _wB\right\rangle _w=k\sum_{j=1}^n\sum_{l=1}^n%
\frac{\partial \overline{A}}{\partial \overline{X_j}}\frac{\partial 
\overline{B}}{\partial \overline{X_l}}\left[ \frac{\partial ^2\overline{S}}{%
\partial \overline{X_j}\partial \overline{X_l}}\right] ^{-1}  \eqnum{10.1}
\end{equation}
where $\overline{A}=\left\langle A\right\rangle _w,\,\left[ a_{jl}\right]
^{-1}$ denotes the inverse of the matrix $a_{jl\text{ }}$and $S=S\left(
X_j\right) $is the entropy of system.

As an example from classical statistical physics we can consider the system
referred in connection with the Eqs. (6.41)-(6.42). The corresponding
stochastic variables are $H$ and $Z_c$. Their fluctuations are described by
the quantities $\Delta _wH,\,\Delta _wZ_c$ and $\left\langle \delta
_wH\delta _wZ_c\right\rangle _w$ whose expressions are given by Eqs. (6.42).

Now we can proceed to a direct examination of the expressions from Eqs.
(6.40), (10.1) and (6.42) of the quantities $\Delta _wA$ and $\left\langle
\delta _wA\delta _wB\right\rangle _w$ which describe the thermal
fluctuations in macroscopic systems. One can observe that all the respective
expressions are structured as products of k with terms which are independent
from k. The alluded independence is ensured by the fact that the mentioned
terms are expressed only by means of macroscopic non-stochastic quantities.
(Note that the mean values $\overline{A}$ from the respective terms must
coincide with deterministic (i.e. nonstochastic) quantities from usual
thermodynamics). Due to the above observed structure, the examined
fluctuation quantities are in a direct dependence of k. So they are
significant respectively negligible as we take $k\neq 0$ or $k\rightarrow 0$%
. Because $k$ is a constant, the limit $k\rightarrow 0$ must be regarded in
the sense that the quantities directly proportional with $k$ are negligible
comparatively with other terms of the same dimensionality but not containing 
$k.$ However, the fluctuations reveal the stochastic characteristics of the
physical systems. Then we can conclude that the thermal stochasticity, for
the system studied in nonquantum statistical physics, is an important
respectively insignificant property as we consider $k\neq 0$ or $%
k\rightarrow 0.$

The mentioned features vis-a-vis the values of $k$ are specific for all the
macroscopic systems (e.g. gases, liquids and solids of various inner
compositions) and for all their specific global variables. But such a remark
reveals the fact that $k$ has the qualities of an authentic {\it generic
indicator for thermal stochasticity} (i.e. for the stochasticity evidenced
through the thermal fluctuations).

Now let us approach questions connected with the {\it quantum stochasticity}
which is specific for the individual, nonrelativistic microparticles of
atomic size. Such a kind of stochasticity is revealed by the specific
quantum fluctuations of the corresponding variables (of orbital and spin
nature). The respective fluctuations described by means of quantities like
the standard deviations and correlations defined in Eqs. (5.2) and (5.15).
Some expressions, e.g. those given by Eqs. (6.19) and (6.38), for the
mentioned fluctuation quantities show the direct dependence of the
respective quantities on the Planck\'{}s constant $\hbar $. Then, there
results that $\hbar $ can play the role of generic indicator for quantum
stochasticity. Corespondingly, as $\hbar \neq 0$ or $\hbar \rightarrow 0$
the mentioned stochasticity appears as a significant respectively negligible
property.

The above mentioned connection between the quantum stochasticity and $\hbar $
must be complemented with certain deeper considerations. Such considerations
regard (Dumitru and Veriest, 1995) different behaviour patterns of various
physical variables in the limit $\hbar \rightarrow 0$, usually called {\bf Q}%
uantum{\it \ }$\rightarrow ${\bf C}lassical{\it \ }{\bf L}imit (QCL).

Firstly, let us refer to the spin variables. We consider an electron whose
spin state is described by the function (spinor) $\chi $ given by 
\begin{equation}
\chi =\left( 
\begin{array}{c}
\cos \alpha \\ 
\sin \alpha
\end{array}
\right) \,\,\,\,\,\,\,\,\,\,\,\,\,\,\,\,\,\,\,\alpha \in \left[ 0,\frac \pi 2%
\right]  \eqnum{10.2}
\end{equation}
For a specific variable we take the z-component of the spin angular moment $%
\widehat{S}_z=\left( \hbar /2\right) \widehat{\sigma _z}$ ($\widehat{\sigma }%
_z$ being the corresponding Pauli matrix). For the respective variable in
the mentioned state we find 
\begin{equation}
\Delta _\chi S_z=\frac \hbar 2\sin 2\alpha  \eqnum{10.3}
\end{equation}
The quantity $\Delta _\chi S_z$ describes the quantum fluctuations of spin
kind i.e. the spin quantum stochasticity. The presence of $\hbar $ in Eq.
(10.3) show that the respective stochasticity is significant or not as $%
\hbar \neq 0$ or $\hbar \rightarrow 0$. This means that $\hbar $ plays the
role of generic indicator for the respective stochasticity. But in the state
described by Eq. (10.2) one finds also $\left\langle S_z\right\rangle
_x=\left( \hbar /2\right) \cos \alpha $. This additional results shows that,
in fact, for $\hbar \rightarrow 0$ the variable $S_z$ disappears completely.
Then we can note that for spin variables the behaviour pattern in quantum $%
\rightarrow $classical limit consists of an annulment of both stochastic
characteristics and mean values (i.e. in a complete disappearance).

In the case of orbital quantum variables the quantum $\rightarrow $classical
limit implies not only the condition $\hbar \rightarrow 0$ but also the
requirement that certain quantum numbers grow unboundedly.. The mentioned
requirement is due to the fact that certain significant variables connected
with the orbital motion (e.g. the energy) pass from their quantum values to
adequate classical values. So, with respect to the mentioned limit the
orbital variables have two kinds of behaviour patterns.

As an example of the first kind we refer to the coordinate $x$ of a harmonic
rectilinear oscillator considered in its n-th energy level. Then similarly
with $\Delta _\Psi \varphi $ from Eq. (6.19) we have: 
\begin{equation}
\Delta _\Psi x=\left[ \frac \hbar {m\omega }\left( n+\frac 12\right) \right]
^{1/2}  \eqnum{10.4}
\end{equation}
where $m,\omega $ and $n$ denote the mass, the angular frequency
respectively the oscillation quantum number. For the mentioned example the
quantum $\rightarrow $classical limit means not only $\hbar \rightarrow 0$
but also $n\rightarrow \infty $. This because the energy must pass from the
quantum expression $E=\hbar \omega \left( n+\frac 12\right) $to the
corresponding classical expression $E_{cl}=\frac 12m\omega ^2x_0^2$, where $%
x_0$ is the coordinate amplitude. Then the standard deviation of $x$ passes
from the quantum value given by Eq. (10.4) to the classical value 
\begin{equation}
\Delta _{cl}x=\frac{x_0}{\sqrt{2}}  \eqnum{10.5}
\end{equation}
But $\Delta _\Psi x$ and $\Delta _{cl}x$ are fluctuation parameters which
describe the stochastic characteristics of $x$ in quantum respectively
classical contexts. Then one can say that in quantum $\rightarrow $classical
limit the above considered coordinate $x$ preserves both its role of
significant variable and its stochasticity.

As an example of the second kind of orbital variable we consider the
distance $r$ between the electron and nucleus in a hydrogen atom. We refer
to an electron in a state described by the wave function $\Psi _{nlm}$ with $%
l=n-1$ (where $n.l$ and $m$ are respectively the principal, orbital and
magnetic quantum numbers). Then for $\Delta _\Psi r=\Delta r$ we can use the
expression given by Schwabl (1995), rewritten in the form 
\begin{equation}
\Delta r=\frac{2\pi \varepsilon _0}{m_0e}\hbar ^2n\left( n+1\right) ^{1/2} 
\eqnum{10.6}
\end{equation}
with $m_0$ and $e$ denoting the mass respectively the change of the
electron. The energy of the electron is 
\begin{equation}
E_n=-\frac{m_0e^4}{32\pi ^2\varepsilon _0^2\hbar ^2n^2}  \eqnum{10.7}
\end{equation}
The quantum$\rightarrow $classical limit requires that $E_n\rightarrow
E_{cl} $ with $E_{cl}$ denoting the classical value of the energy. Then from
Eqs. (10.6) and (10.7) it results that in the respective limit we have 
\begin{equation}
\Delta r\rightarrow \left( \frac{\hbar e^4}{16\pi \varepsilon _0}\right)
^{1/2}\left( -2m_0E_{cl}\right) ^{-1/4}  \eqnum{10.8}
\end{equation}
In the same circumstances we obtain 
\begin{equation}
\left\langle r\right\rangle _\Psi \rightarrow r_{cl}=-\frac{e^2}{8\pi
\varepsilon _0E_{cl}}  \eqnum{10.9}
\end{equation}
So it results that in the quantum$\rightarrow $classical limit (when $\hbar
\rightarrow 0$ and $E_n\rightarrow E_{cl}$) we have $\Delta r\rightarrow 0$
and $\left\langle r\right\rangle _\Psi \rightarrow r_{cl}\neq 0.$ This means
that $r$ preserves its role of significant variable but loses its
stochasticity.

The above considerations can be concluded with the following remark:

\underline{${\bf P-10.1}$} : In the quantum$\rightarrow $classical limit the
physical variables display the following different behaviour patterns:

(i) The complete disappearance of both stochastic characteristics and mean
values, as in the case of spin variables.

(ii) The preservation of both the role of significant variable and of
stochastic characteristics, as in the case of oscillator coordinate

(iii) The preservation of the role of significant variable but the loss of
stochastic characteristics as in the case of electron-nucleus distance.$%
\blacktriangle $

It is clear that the above remark corrects the traditional belief of a
unique behaviour pattern compulsorily associated with the disappearance of
the ''uncertainties'' (i.e. of the standard deviations $\Delta _\Psi A$).

Now let us return to the quantum stochasticity, specific for the variables
of individual microparticles. We think that, in spite of the peculiarities
mentioned in ${\bf P-10.1}$, the Planck constant $\hbar $ can be considered
as a generic indicator of such a stochasticity. Moreover, we consider that
the respective role of $\hbar $ is completely similar with that the
Boltzmann constant k with respect to the macroscopic thermal stochasticity
(see above).

Regarding the mentioned roles of $\hbar $ and $k$ another observation must
be added. In the discussed cases, $\hbar $ and $k$ appear independently and
singly. That is why one can say that the stochasticity of the corresponding
systems (microparticles and classical macroscopic systems) has a onefold
character. But there are also physical systems endowed with a twofold
stochasticity characterized by a simultaneous connection with both $\hbar $
and $k$. Such systems are those studied in quantum statistical physics,
i.e., the bodies of macroscopic size considered as statistical ensembles of
quantum microparticles. The stochasticity of the respective systems is
revealed by corresponding fluctuations described by the quantities given by
Eqs. (5.24) which depend simultaneously on both $\hbar $ and $k$. The
respective dependence is revealed by the so-called fluctuation-dissipation
theorem. According to the respective theorem (Kubo 1957; Zubarev 1971;
Balescu 1975) one can write 
\[
\left\langle \delta _\rho \widehat{A}\delta _\rho \widehat{B}\right\rangle
_\rho +\left\langle \delta _\rho \widehat{B}\delta _\rho \widehat{A}%
\right\rangle _\rho = 
\]
\begin{equation}
=\frac i{2\pi }\int_{-\infty }^\infty \hbar \coth \left( \frac{\hbar \omega 
}{2kT}\right) \left[ \chi _{AB}^{*}\left( \omega \right) -\chi _{BA}\left(
\omega \right) \right] d\omega  \eqnum{10.11}
\end{equation}
with $\chi _{AB}^{*}\left( \omega \right) $ as complex conjugate of $\chi
_{AB}\left( \omega \right) .$

In Eq. (10.11) $\chi _{AB}\left( \omega \right) $ represent the generalized
susceptibilities which appears also in the deterministic framework of
nonequilibrium thermodynamics (De Groot and Mazur, 1962). But it is a known
fact that in the respective framework all the stochastic characteristics of
physical variables are neglected and no miscroscopic (i.e. atomic or
molecular) structure of the systems is taken into account. Another fact is
that $\chi _{AB}\left( \omega \right) $ are directly connected (Landau and
Lifschitz, 1984) with the macroscopic nonstochastic expression of the energy
dissipated inside the thermodynamic systems actioned by external
deterministic and macroscopic perturbations. The mentioned facts show that
the susceptibilities $\chi _{AB}\left( \omega \right) $ do not depend on the
constants $\hbar $ and $k.$

The above mentioned property of $\chi _{AB}\left( \omega \right) $ combined
with Eq. (10.11) shows that the sole significant dependence of fluctuation
quantities given by Eq. (5.24) on the constants $k$ and $\hbar $ is given by
the factor $\hbar \coth \left( \hbar \omega /kT\right) $. For the respective
factor one can write 
\begin{equation}
\lim_{k\rightarrow 0}\left\{ \lim_{\hbar \rightarrow 0}\left[ \hbar \coth
\left( \frac{\hbar \omega }{2kT}\right) \right] \right\} =0  \eqnum{10.12}
\end{equation}
This means that when both $\hbar $ and $k$ tend to zero the fluctuation
quantities defined by Eq. (5.24) become null. So it results that in the
mentioned limit the fluctuations in quantum statistical systems cease to
manifest themselves. Consequently, for such a limit the respective systems
lose their stochastic characteristics. Then in the spirit of the above
presented opinions one can state that quantum statistical systems can be
considered as endowed with a twofold stochasticity of the thermal and
quantum type, revealed respectively by $k$ and $\hbar $ as generic
indicators.

In the above considerations the stochasticity appears as a property of
exclusively intrinsic type. This means that it is connected only with the
internal (inner) characteristics of the considered systems and does not
depend on external (outside) factors. Moreover the mentioned stochasticity
is directly and strongly associated with $\hbar $ and $k$ as generic
indicators. But the stochasticity can be also of extrinsic type. In such
cases it is essentially connected with factors from the outside
(surroundings) of the considered physical systems. Also the extrinsic
stochasticity is not (necessarily) associated with $\hbar $ and $k.$ As
examples of system with stochasticity of exclusively extrinsic type can be
considered an empty bottle floating on a stormy sea or a die in a game.

In practice one finds also systems endowed with stochasticity of both the
intrinsic and extrinsic types. Such are for example, the electric and
electronic circuits. For a circuit the intrinsic stochasticity is caused by
the thermal agitation of the charge carries and/or of elementary (electric
or magnetic) dipoles inside of its constitutive elements (i.e. inside of
resistors, inductances, condensers, transistors, integrated circuits, etc.).
The respective agitation is responsible for fluctuations of macroscopic
voltages and currents. Such fluctuations are known (Robinson, 1974) as
thermal (or Nyquist) noises. Note that in the case of circuits the intrinsic
stochasticity is characterized by generic indicators. Such indicators are $k$
alone, if the circuit is considered as a classical (nonquantum) statistical
system, and $k$ together with $\hbar $ when the circuit is viewed as a
quantum statistical system. Otherwise, the stochasticity of a circuit can
also be of extrinsic type when it is under the influence of a large variety
of factors. Such factors can be: thermal fluctuations in the surrounding
medium, accidental outside discharges and inductions, atmospheric (or even
cosmic) electrical phenomena. The mentioned extrinsic stochasticity is also
responsible for the noises in macroscopic currents and voltages in the
circuits. But it must be noted that even for circuits, the extrinsic
stochasticity is not connected in principle with generic indicators
dependent on fundamental physical constants (such as $\hbar $ and $k$). As
an interesting case which implies stochasticity of both intrinsic and
extrinsic type can be considered a measuring process viewed as in Sec. IX.
In such a case the intrinsic stochasticity regards the inner properties of
the measured system while the extrinsic one is due to the measuring
apparatus. The corresponding intrinsic stochasticity is connected with $%
\hbar $ and $k$ as generic indicators in the above discussed sense. However,
the extrinsic stochasticity, due to the apparatus, seems to be not connected
with certain generic indicators. This because of the large diversity of
apparata as regards their own structure and accuracy.

\section{A FEW\ REMARKS\ ON\ SEVERAL\ ADJACENT\ QUESRTIONS}

With respect to problematics of HR proper, in literature (see the
bibliographical publications mentioned in the end of Sec. I) one knows of a
large variety of adjacent questions which, one way or another, are allied
with the subjects discussed in the previous sections of this paper. Now, we
wish to note a few remarks on several such questions.

$\blacklozenge $ Firstly, let us refer to the consequences of the
here-proposed reconsideration of HR for both lucrative procedures and
interpretational frame of quantum mechanics. Note that our reconsideration
does not prejudice in any way the authentic version of the mentioned
lucrative procedures, which, in fact, have met with unquestionable successes
in both basic and aplicative researches. As regards the alluded
interpretational frame, our reconsideration, mainly by the abandonment of
TIHR, generates major and important changes. But such changes must be
regarded as benefic, since they can offer a genuine elucidation to the
controversial questions introduced in the frame of science by TIHR.$%
\blacksquare $

$\blacklozenge $ By reinterpreting HR in the sense presented in Sec.VIII the
respective relations lose their quality of crucial physical formulae. So,
one can find a consonance with the prediction (Dirac, 1963): ''I think one
can make a safe guess that uncertainty relations in their present form will
not survive in the physics of future''. Note that the above prediction was
founded not on some considerations about the essence of HR but on a
supposition about the future role of $\hbar $ in physics. So it was supposed
that $\hbar $ will be a derived quantity while $c$ and $e$ (speed of light
and elementary charge) will remain as fundamental constants. That is why we
wish to add here that our view about HR does not affect the actual position
of $\hbar $ as a physical constant. More precisely, our findings cannot
answer the question whether $\hbar $ will be a fundamental constant or a
derived quantity (e.g. expressed in terms of $c$ and $e$).$\blacksquare $

\smallskip $\blacklozenge $ As it was pointed in Sec.X the Planck's constant 
$\hbar \smallskip $ has also the significance of estimator for the spin of
microparticles (like the electron). So the spin appears to be a notable
respectively absent property as $\hbar \neq 0$ or $\hbar \rightarrow 0$. On
the other hand with the reference to the spin there are also some intriguing
questions related to its relativistic justification. Usually (Dirac, 1958;
Blochintsev,1981) for electrons the spin is regarded to be essentially
explicable as consequence of relativistic theory. But, as it is known, the
relativistic characteristics of a particle are evidenced by the relative
value $v/c$ of its velocity $v$ comparatively with the light velocity $c$.
Particularly, the respective characteristics must be insignificant when $%
v/c\ll 1$ or $c\rightarrow \infty $. Then the absence of the factor $v/c$
(or of some other equivalent factors) in the description of the electron
spin variable appears at least as intriguing fact. Is such a fact a
sufficient reason to consider $\hbar $ as a derived quantity in the sense
guessed by Dirac (1963). In such a sense $\hbar =\frac{e^2}{4\pi \cdot
\varepsilon _0\cdot c\cdot \alpha }$ ($\varepsilon _o$= the permittivity of
vacuum and $\alpha =\frac 1{137}$ = the fine structure constant) and the
situations with $\hbar \rightarrow 0$ appear when $c\rightarrow \infty $. So
the significance of $\hbar $ as spin estimator can be apparently related
with some aspects of relativity. But here it must be noted the surprising
fact that even in the nonrelativistic {}limit (i.e. when $v/c\ll 1$ or $%
c\rightarrow \infty $) the spin remains a significant variable of the
electron. It is known (Ivanov, 1989) that the electron spin plays a decisive
role (as a fourth quantum variable/number) in the electronic configuration
of many-electronic atoms, in spite of the fact that for atomic electrons $%
v/c\ll 1$. Due to the here mentioned features we think that the relativistic
justification of the spin appears as a intriguing question witch requires
further investigations.$\blacksquare $

$\blacklozenge $ Our findings facilitate also a remark in connection with
another supposition about $\hbar $. The respective supposition regards the
possible existence of multiple Planck constants associated with various
kinds of microparticles (e.g. with electrons, protons, neutrons). Currently,
(Whichman, 1971; Fischbach {\it et. al.} 1991), the tendency is to contest
such a possibility and to promote the idea of a unique Planck constant. For
this one appeals either to experimental data or to some connection with the
fundamental conservation laws. We think that our view about $\hbar $ pleads
somewhat for the alluded idea of uniqueness.. So, regarding $\hbar $ as
generic indicator of quantum stochasticity, this one must have the same
value for various kinds of microparticles. This because, similarly, the
Boltzman constant $k$ in its role of generic indicator for thermal
stochasticity has a unique value for various kinds of macroscopic systems
(e.g. hydrogen gas, liquid water or crystallin germanium ).

$\blacklozenge $ The revealed stochastic similarity among quantum
microparticles and macroscopic systems facilitates another remark. In the
macroscopic case the stochastic characteristics for {\it an individual system%
} is incorporated in the probability distribution $w$ (see sections V and
VI. K). As we have shown, the quantum similar of $w(x)$ is the wave function 
$\Psi $ (or the square $\left| \Psi \right| ^2$ of its module). Such a $%
w-\Psi $ similarity motivates us to agree the idea (Van Kampen 1978) that $%
\Psi $ refer to a single system (microparticle). Simultaneously, we incline
to a circumspect regard about the opinions that $\Psi $ belongs to an
''ensemble of equally prepared systems'' (Tschudi, 1987) or to an ''abstract
physical object'' (Mayants,1984). Moreover, our agreement and opinion are
also motivated by the observation that in practical applications both $\Psi $
and $w$ are calculated for individual systems (e.g. for an electron in a
hydrogen atom or, respectively for an ideal gas).$\blacksquare $

$\blacklozenge $A distinct group of remarks regards the reduction of
stochasticity to subjacent elements of deterministic nature, for both cases
of thermodynamic systems and quantum microparticles. In the first case the
stochasticity refers to the macroscopic variables which characterize each
system as a whole. But according to the classical statistical mechanics the
respective variables are expressible in terms of subjacent molecular
quantities (coordinates and momenta) which are considered as deterministic
elements. In the case of quantum microparticles a similar problem was taken
into account. So it was promoted the idea that the stochastic quantum
variables (characterizing each microparticle as a whole) would be
expressible in terms of some sujacent elements of deterministic nature,
called ''hidden variables''. Viewing comparatively the two mentione d cases
we think that is of nontrivial interest ti note the foolowing observations:

(i) In the case of thermodynamic systems the subjacent molecular quantities
can be justfied in essence only by adequate experimental facts.

(ii) The mentioned molecular quantities are deterministic (i. e.
depresionfree) only from a microscopic perspective, connected with the
characteristics of the molecules. From a macroscopic perspective, coonected
with a thermodynamic systems as a whole, they are stochastic variables. That
is why, for example, in respect with a thermodynamic system like an ideal
gas one speaks about the mean value and non-null dispersion of the molecular
velocity.

(iii) Even by taking into account the existence of sujacent molecular
quantities the macroscopic variables, characterizing a thermodynamic system
as a whole, keep their stochastic characteristics. Particularly the
mentioned existence does not influence the verity or the significance of the
macroscopic relations frim the family of Eqs. (5.21) - (5.23).

(iv) The above observations (i) - (iii) reveal as unfounded the idea (uffink
and Van Lith 1999) that the sole examination of some theoretical formulas,
from the mentioned family, can give a light on the problem of reduction of
thermodynamic stochasticity to subjacend deterministic elements.

(v) By analogy with the fact noted in (i), in the case of quantum
microparticles, the existence of the ''hidden variables'' must be proved
firstly by indubitable experimental facts. But, as far as we know, until now
such an experimental proof was not ratified by scientific research.

(vi) The existence of the mentioned ''hidden variables'' cannot be asserted
only by means of considerations on some theoretical formulas regarding the
global stochasticity of quantum microparticles, such are the HR.

(vii) The global description of a quantum microparticle remain equally
probabilistic in both cases, with or without ''hidden variables''. More
exactly in both casses for a variable refering to a quantum microparticle as
a whole the theoretical predictions must be done in probabilistic terms
while the experimental informations can be obtained only from measurements
consisting in statistical samplings.

$\blacklozenge $ The discussions from Sec. X about the stochasticity suggest
a remark connected with the Boltzmann\'{}s constant $k$. As we have shown $k$
plays a major role in the evaluation of the level of the thermal
stochasticity. But the respective stochasticity must be regarded as an
important property of the macroscopic systems. So one finds as unfounded the
idea, promoted in some publications (Wichman, 1971; Landau and Lifschitz,
1984; Storm, 1986) that, in physics $k$ has only a minor role of conversion
factor between temperature scales (from energetic units into Kelvin degrees).%
$\blacksquare $

\pagebreak

\section{CONCLUSIONS}

We started the paper reminding the fact that even in our days TIHR persists
as a source of unelucidated controversies about it defects. Motivated by the
respective fact we proposed an investigation in the very core of the alluded
controversies and defects. For such a purpose firstly we identified the main
elements (assertions and arguments) of TIHR. Then, in reference with the
mentioned elements, we localized the most known and critical defects of TIHR.

In such a reference frame we analyzed the reality of the respective defects.
We found that all of them are veridical. Moreover, for TIHR, they are
insurmountable and incriminate each of its main elements. So we can conclude
that the sole reasonable attitude is to abandon TIHR as an unjustified
doctrine.

The mentioned abandonment must be accompanied with a search for a new and
genuine {}reinterpretation of HR. On this direction we opine that HR of
troughs-experimental nature must be disregarded because they are fictitious
formulae without a true physical significance. On the other hand we think
that the theoretical HR are authentic physical formulae regarding the
quantum fluctuations. So regarded the theoretical HR belong to a large class
of formulae specific for systems, of both quantum and non-quantum nature,
endowed with stochastic characteristics. By adopting the mentioned regards
about HR all the controversies connected with the TIHR are elucidated on a
natural way.

In the mentioned regard HR lose their traditional role of crucial physical
formulae connected with the description of measurement characteristics
(uncertainties). In the here promoted view the respective description must
be done in terms (and formulae) which do not belong to the traditional
chapter of physics (including the quantum mechanics). We suggested that a
promising version for the description of measurements can be done in terms
of information theory. So a measurement can be considered as an information
transmission, from the measured system (information source) through the
measuring device (transmission channel) to the device recorder (information
receiver). Then the measuring {}uncertainties appear as alternations of
processed information. In the Sec. IX we illustrated the alluded
informational model with some concrete considerations.

In our opinion the theoretical HR and their classical (non-quantum) similars
are connected with the stochasticity regarded as an important property of
physical systems. We showed that the respective property is characterized by
generic indicators which are:

(i) the Planck's constant $\hbar $ (for quantum microparticles),

(ii) the Boltzmann's constant $k$ (for classical thermodynamical system),
respectively

(iii) both $\hbar $ and $k$ (for quantum statistical systems).

In the end, in Sec. XI, we presented remarks on some questions which are
adjacent with the subjects discussed in the other parts of the paper.

\section*{ACKNOWLEDGMENTS}

$\blacklozenge $ Several publications studied by me in connection with this
and other previous papers of mine were put at my disposal by their authors
(often in a preprint or amended-reprint form). To all the respective authors
I express my sincere thanks.

$\blacklozenge $ Refering to my own views, during the years, I have received
many comments which stimulated my work. I remain profoundly grateful to the
corresponding commentators (referees and readers of my papers, colleagues).
But, of course, that for all the shortcomings of my views I assume the
entire responsibility.

$\blacklozenge $ The work involved in the long-standing studies reported
here took me away from some family duties. For the evinced agreement as well
as for the permanent aid I am deeply indebted to all my family.

$\blacklozenge $ In the end I mention that the research reported here was
finalized with a partial support from the Roumanian Ministry of National
Education under a grant.

\pagebreak

\section*{LIST OF ACRONYMS}

\begin{tabbing}

CRCRCRCRCR \= correlation relation(s) \kill\\

CR\>  correlation relation(s)\\

HR\> Heisenberg's relation(s)\\

P\>  point\\

P.../A\> assertion point\\

P.../M\> motivation point\\

QCL\> quantum$\longrightarrow $classical limit\\

QTP\> quantum torsion pendulum\\

SRTE\> super-resolution thought experiment(s/al)\\

TE\> thought experiment(s/al)\\

TIHR\> traditional interpretation of Heisenberg's relations\\

\end{tabbing}

\begin{figure}
\caption{ Private paper from J.S. Bell to the author (dated January 29,1985) }
\label{}
\end{figure}

\end{document}